\documentclass[preprint2]{aastex}

\begin{document}

\bibliographystyle{apj}

\title{A Unified Catalog of Radio Objects Detected by NVSS, FIRST, WENSS, GB6,
  and SDSS} 

\author{Amy E. Kimball\altaffilmark{1}}
\author{\v{Z}eljko Ivezi\'{c}\altaffilmark{1}}

\altaffiltext{1}{Department of Astronomy, University of Washington, Box 351580,
  Seattle, WA 98195, USA; akimball@astro.washington.edu,
  ivezic@astro.washington.edu}

\begin{abstract} 
We construct a catalog of radio sources detected by GB6 (6\,cm), FIRST and NVSS
(20\,cm), and WENSS (92\,cm) radio surveys, and the SDSS optical survey.  The
2.7\,million entries in the 
publicly-available master catalog are comprised of the closest three FIRST to
NVSS matches (within 30\arcsec) and vice-versa, and unmatched sources from each 
survey.  Entries are supplemented by data from the other radio and optical
surveys, where available.  All objects with even a small probability of
physical association are included, such that catalog users can easily implement
their own selection criteria for data analysis.  We perform data analysis in
the $\sim3000\deg^2$ region of sky where the surveys overlap, which contains 
140,000 NVSS-FIRST sources, of which 64,000 are detected by WENSS and 12,000 by
GB6.  About one third of each sample is detected by SDSS.  An automated
classification method based on 20\,cm 
fluxes defines three radio morphology classes: complex, resolved, and compact.
Radio color-magnitude-morphology diagrams for these classes show structure 
suggestive of strong underlying physical correlations.  Complex and
resolved sources tend to have a steep spectral slope ($\alpha\sim-0.8$) that is 
nearly constant from 6 to 92\,cm, while the compact class (unresolved on
$\sim5\arcsec$ scale by FIRST) contains a significant number of flat-spectrum
($\alpha\sim0$) sources.  In the optically-detected sample, quasars dominate
the flat-spectrum compact sources while steep-spectrum and resolved objects
contain substantial numbers of both quasars and galaxies.  Differential radio
counts of quasars and galaxies are similar at bright flux levels ($>100$\,mJy
at 20\,cm), while at fainter levels the quasar counts are significantly reduced
below galaxy counts.  The optically-undetected sample is
strongly biased toward steep-spectrum sources.  In samples of quasars and
galaxies with SDSS spectra (2,885 and 1,288 respectively), we find that radio
properties such as spectral slope, morphology, and radio loudness are
correlated with optical color and luminosity.

\end{abstract}


\section{INTRODUCTION}
\label{sec:introduction}

Quasars and powerful radio galaxies dominate the observed counts of continuum
radio sources above milliJansky flux levels, and display spectacular
morphological variety that is correlated with other properties such as spectral
slope and luminosity.  The unification paradigm for radio galaxies and quasars
\citep{urry_review,jackson99}
attempts to explain much of this rich variety of observational data as arising
from essentially the same anisotropic processes which appear very different to
us because of varying viewing angles to the radio jets.  This conjecture has
fundamental implications for our understanding of quasars and galaxies, but for
a strong test of the unification paradigm one essentially needs a large
statistical sample with well-controlled selection criteria and robust estimates
of the source morphology, as well as appropriate models to interpret the data. 

Statistical studies of radio emission from extragalactic sources are entering a
new era, resulting from the availability of large sky radio surveys that are
sensitive to milliJansky flux levels \citep[e.g.,][]{first,nvss,debreuck}.  The
catalogs based on these surveys contain large numbers of sources, have high
completeness and low contamination, and are available in digital form.  The
wide wavelength region spanned by these surveys, from 6\,cm for GB6 to 92\,cm
for WENSS, and detailed morphological information at 20\,cm provided by FIRST
and NVSS, allow significant quantitative and qualitative advances in studies of
radio sources.  In addition, the optical catalog obtained by the Sloan Digital
Sky Survey \citep[SDSS;][]{york} can be used to separate quasars from galaxies,
and the redshifts measured by SDSS allow a comprehensive study of the
optical-radio correlation for quasars and galaxies. 

FIRST and NVSS, conducted separately
at the Very Large Array (VLA), were the first radio surveys with sufficiently
high angular resolution to allow unambiguous matching with deep optical
surveys, providing identifications for a large number of radio sources.
The two surveys have the same radio frequency, but FIRST goes
slightly deeper with higher resolution and smaller sky coverage
(\S\ref{sec:surveys}). 
\citet{machalski} and \citet{sadler99,sadler02} measured the radio luminosity
function of radio-loud active galactic nuclei (AGN) and star-forming
galaxies by  cross-correlating the NVSS with spectroscopic galaxy surveys.
\citet{magliocchetti} matched FIRST to the 2 degree Field Galaxy Redshift Survey
\citep{2dfgrs}, using spectra to classify galaxies as ``classical'' radio
galaxies, starburst and late-type galaxies, and Seyfert galaxies.  \citet{i02}
cross-correlated the FIRST survey with the SDSS photometric survey, resulting
in a much larger sample of optical identifications (about one third of FIRST
sources are matched to an SDSS source, and 0.16\% of SDSS sources are matched
to a FIRST source, with a matching radius of $2\arcsec$), with galaxies 
outnumbering quasars 5:1, and a negligible fraction of radio stars in the
sample.  \citet{best} cross-correlated the SDSS spectroscopic galaxies sample
with both FIRST and NVSS, and found that AGN dominate radio counts down to
5\,mJy at 20\,cm.  The more recent work of \citet{mauch07} includes the largest
radio-selected galaxy sample available from a single radio survey, combining
the NVSS with the 6 degree Field Galaxy Survey \citep{6dfgs}; they confirm that
radio-loud AGN and star-forming galaxies have quite different distributions in 
the plane of radio power versus absolute K band (infrared) magnitude.
The study of radio and optical properties of
quasars was extended by \citet{devriesW} who found that 10\% of SDSS quasars
have detectable radio cores at 1.4\,GHz ($>0.75$\,mJy), and 1.7\% have
double-lobed morphology, i.e., are associated with multiple FIRST components.

In this paper, we describe the construction of a unified catalog 
of radio sources detected in the 6\,cm GB6, 20\,cm FIRST and NVSS, and 92\,cm
WENSS radio surveys, and the SDSS optical survey (Data Release 6, hereafter
DR6).  We began by merging the FIRST and NVSS surveys into 
a single catalog containing over 2\,million sources detected by at least one
survey; about 500,000 sources are detected by both FIRST and NVSS.  Where
available, we supplement this survey with data obtained by the GB6 and WENSS
surveys that enable the computation of radio spectral slopes; about 30,000
sources are detected by all four radio surveys.  The radio sources were also
cross-correlated with the optical SDSS catalogs; nearly 92,000 FIRST
sources have an SDSS counterpart within $2\arcsec$.

The radio and optical surveys chosen for inclusion in the unified catalog
primarily cover the northern celestial hemisphere.  We have opted to use
these surveys in order to take advantage of the high astrometric accuracy of
FIRST, which, designed to cover the same region of sky as the SDSS, is limited
to the northern Galactic cap.  Several recent
large-area radio surveys covering a wide range of frequencies are also
available in the south.  These include the Sydney University Molonglo Sky
Survey at 36\,cm \citep[SUMSS;][]{mauch03}, the
Parkes-MIT-NRAO survey at 6\,cm \citep[PMN;][]{gregory}, and the Australia
Telescope 20-GHz (1.5\,cm) survey \citep[AT20G;][]{massardi}.

The main advantages of the unified catalog presented in this paper are the
multi-wavelength radio data (92\,cm and 6\,cm in addition to 20\,cm), and the
increased number of optical identifications from the SDSS DR6.  
By combining these five extensive surveys, we have assembled precise
astrometric measurements, flux at multiple wavelengths, spectral index,
morphological 
information (size, shape, and orientation of resolved objects), and optical
identifications into a single comprehensive catalog of radio objects.  At
optical wavelengths, the SDSS provides colors and classification (i.e., into
quasars and galaxies) for the $\approx30\%$ of sources detected optically.  The
unified catalog is primarily a resource of low-redshift ($z\lesssim2$) quasars
and radio galaxies with AGN, but may also prove to be an important source of
rarer objects such as radio stars, high-redshift quasars, and high-redshift
galaxies.  The area observed by all five surveys is nearly $3,000\deg^2$.  The
unified catalog provides comprehensive multi-wavelength observations at greater
depth and for a larger number of sources than any previously available catalog
of radio objects.

\begin{figure}
\figcaption{\label{fig:position plots}
Using sparse sampling, we show the sky coverage of each of the 
surveys used to created the matched catalog.  The blue line shows the 
$2955\deg^2$ region of overlap of the radio surveys and the SDSS photometric
survey.  The $2894\deg^2$ region outlined by the orange and blue lines shows
the region included in the spectroscopic sample of SDSS DR6.  The dashed red
line indicates the Galactic plane.}
\epsscale{0.8}
\plotone{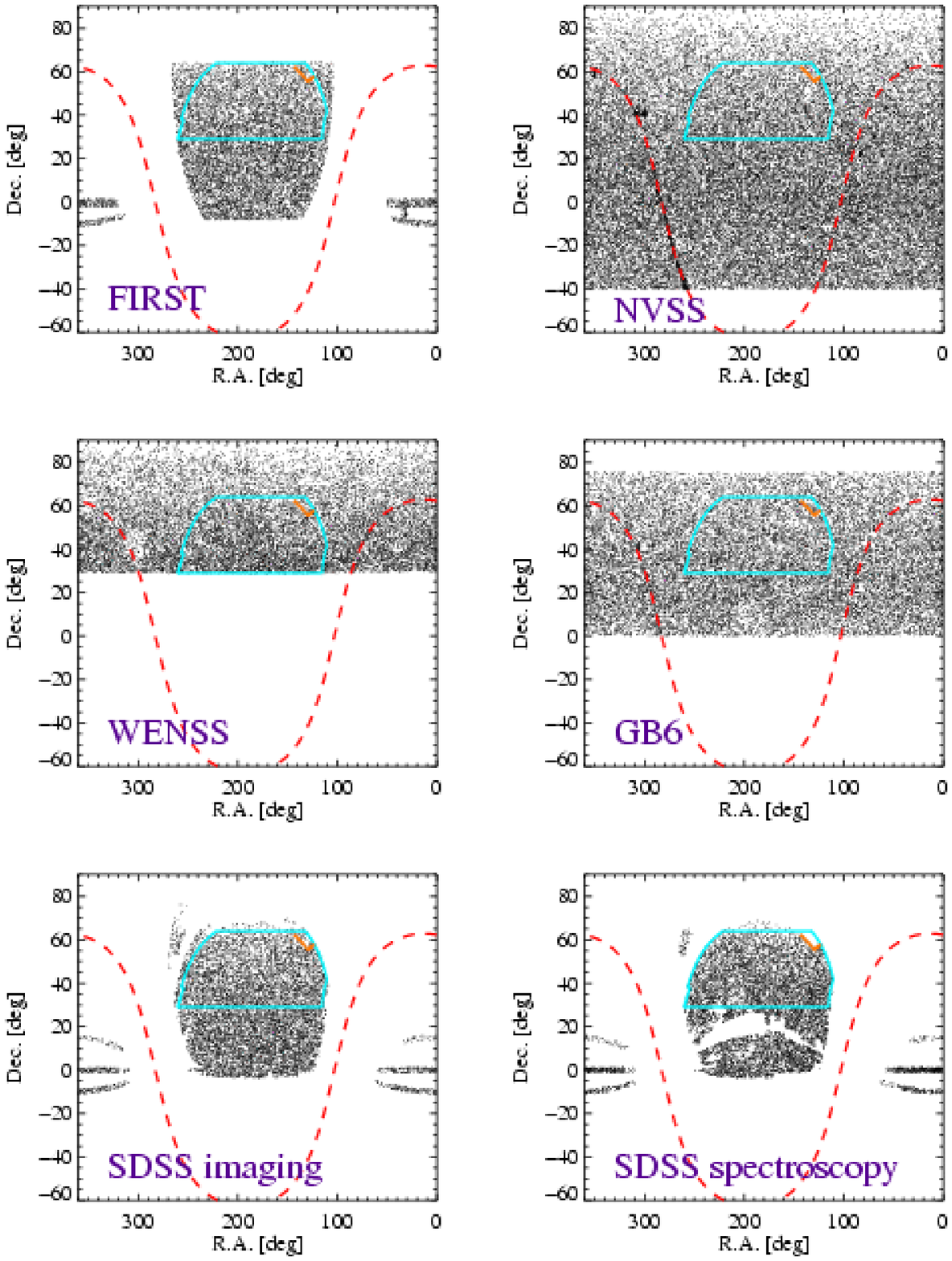}
\end{figure}

The limits in sky coverage of the catalog are defined by the FIRST and NVSS sky
coverage (Fig.~\ref{fig:position plots}).  The unified catalog 
thus includes sources in the Galactic plane which are imaged by the NVSS.
Galactic sources must be studied with care: due to the nature of
interferometry, radio surveys (depending on their angular resolution) generally
do a poor job of imaging highly-extended sources such as Galactic HII regions
and supernova remnants, whose angular sizes often reach several arcmin or
more.  The analysis presented in this paper is limited to the sky covered by 
FIRST, which is greater than $30\degr$ from the Galactic plane.

In this paper, we discuss scientific applications of the unified radio catalog
and present a preliminary data analysis.  In a companion paper (A. Kimball et
al. in preparation), we will expand
upon and refine this analysis by comparing  the source distribution in radio
morphology, radio color, and optical classification space to the models of
\citet{barai06,barai07}.

The remainder of the paper is laid out as follows.
In \S\ref{sec:surveys}, we describe the surveys used to create the
unified radio catalog.  In \S\ref{sec:matching} we discuss the creation of the
catalog, including completeness and efficiency of the matching
algorithms.  In \S\ref{sec:analysis} we present a preliminary analysis of the
radio source distribution according to radio morphology, radio color, optical
identification, and, for sources with spectra, redshift and luminosity.  We also
discuss some catalog applications.  We summarize our results, and discuss the
suitability of the catalog for comparison with radio evolution models,
in~\S\ref{sec:discussion}.

\section{SOURCE SURVEYS}
\label{sec:surveys}

Before proceeding with a description of merging procedure for the FIRST and NVSS
catalogs, we briefly describe each survey used in creating the multi-wavelength
radio catalog, including the sky coverage, wavelength, astrometric accuracy, and 
flux limit (see Table~\ref{table:surveys} for summary). Sky coverage of each
survey is shown in Figure~\ref{fig:position plots}.  The region of sky observed
by all of the contributing surveys is indicated by the cyan solid line.

\subsection{Radio surveys}

\subsubsection{FIRST}
\label{subsubsec:first}

The FIRST survey \citep[Faint Images of the Radio Sky at Twenty
  centimeters;][]{first} used the VLA to observe the sky at
20\,cm (1.4\,GHz) with a beam size of 5\farcs4 and an rms sensitivity of about
0.15\,mJy~beam$^{-1}$.  Designed to cover the same region of the sky as the SDSS,
FIRST observed $9,000\deg^2$ at the north Galactic cap and a smaller
$\sim2.5\degr$ wide strip along the Celestial Equator.  It is 95\% complete to
2\,mJy and 80\% complete to the survey limit of 1\,mJy.  The survey contains
over 800,000 unique sources, with astrometric uncertainty of
$\lesssim1\arcsec$.  FIRST includes two measures of 20\,cm continuum flux
density: the peak value, $F_\mathrm{peak}$, and the integrated flux density, 
$F_\mathrm{int}$, measured by fitting a two-dimensional Gaussian to co-added
images of each source.

\subsubsection{NVSS}

The NVSS \citep[NRAO-VLA Sky Survey;][]{nvss} was also carried out using the
VLA radio telescope to observe the sky at 20\,cm (1.4\,GHz), the same
wavelength as FIRST. However, the NVSS used a different antenna configuration,
resulting in a lower spatial resolution ($45\arcsec$~beam$^{-1}$).  Lower
resolution radio surveys provide more accurate flux measurements for extended 
sources, where high-resolution surveys can miss a significant fraction of the
flux. The astrometric accuracy ranges from $1\arcsec$ for the brightest NVSS
detections to about $7\arcsec$ for the faintest detections.  The survey covers
the entire sky north of $\delta=-40\degr$ and contains over 1.8\,million unique
detections brighter than 2.5\,mJy.

For the analysis of this paper, we adopt integrated NVSS flux densities rather
than the peak flux densities reported in the NVSS catalog.  \citet{nvss}
provides formulas for converting peak flux densities to integrated flux
densities, as well as formulas for errors in NVSS measured and calculated
values.  The unified catalog includes the integrated NVSS flux density, the
corrected peak flux density, and the deconvolved major and minor axis sizes.

\subsubsection{WENSS}

WENSS \citep[Westerbork Northern Sky Survey;][]{wenss} is a 92\,cm (325\,MHz)
survey completed with the Westerbork Synthesis Radio Telescope.
It maps the radio sky north of $\delta=29\degr$ to a limiting flux
of 18\,mJy, with a beam size of
$54\arcsec\times54\arcsec\mathrm{cosec}(\delta)$.  The complete survey contains
almost 220,000 sources, with a positional uncertainty of $\lesssim1.5\arcsec$
for bright sources and $\lesssim5\arcsec$ for faint sources. 

\subsubsection{GB6}

The GB6 survey at 4.85\,GHz \citep[Green Bank 6\,cm survey;][]{gb6} was
executed with the (now defunct) 91m Green Bank telescope in 1986 November and
1987 October.  Data from both epochs were assembled into a survey covering the
$0\degr<\delta<75\degr$ sky down to a limiting flux of 18\,mJy, with
$3.5\arcmin$ resolution.  GB6 contains
over 75,000 sources, and has a positional uncertainty of about $10\arcsec$
at the bright end and about $50\arcsec$ for faint sources.

\subsection{Optical Surveys: SDSS}

We use photometric and spectroscopic coverage from the sixth data
release (DR6) of the Sloan Digital Sky Survey \citep[SDSS; see][and references
  therein]{york,edr,dr5}.  The survey, not yet finished but nearing completion,
will eventually cover $10,000\deg^2$ in the northern galactic cap and a smaller
region on the celestial equator.  DR6 covers roughly $9,600\deg^2$, and
contains photometric observations for 287\,million unique objects, as well as
spectra for more than 1\,million sources. 

\subsubsection{The SDSS Photometric Survey}
\label{subsubsec:sdss photometry}

The SDSS photometric survey contains flux densities of detected objects
measured nearly simultaneously in five wavelength bands \citep[$u$, $g$, $r$,
  $i$, and $z$;][]{fukugita} with effective wavelengths of 3551, 4686, 6165,
7481, and 8931\AA~\citep{gunn}.  The catalog is 95\% complete for point sources
to limiting AB magnitudes of 22.0, 22.2, 22.2, 21.3, and 20.5 respectively.
Typical seeing is about $1.4\arcsec$ and positional uncertainty is less than
about $0.1\arcsec$ \citep[rms per
  coordinate for sources with $r<20.5$,][]{sdss_astrometry}.  
The photometry is repeatable to 0.02
mag \citep[rms for sources not limited by photon statistics,][]{ivezic03} and
with a zeropoint uncertainty of $\sim$0.02-0.03 mag \citep{ivezic04}.  A
compendium of other technical details about SDSS can be found on the SDSS web
site\footnote{\emph{http://www.sdss.org}}, which also provides an interface for
public data access.

Optical magnitudes in this paper refer to SDSS \emph{model} magnitudes,
measured using a weighting function determined from the object's $r$ band
image \citep[see][]{edr}.  The weighting function represents the better fitting
of an exponential profile and a de Vaucouleurs profile.  The chosen model is
then used to determine the magnitude in all five bands.  We corrected all
magnitudes for Galactic extinction following \citet{sfd}.  When selecting
candidate matches from the SDSS, we required unique sources that are brighter
than $r=22.2$ or brighter than $z=21.2$.

The morphological information from SDSS images allows reliable star-galaxy
separation to $r\sim21.5$ \citep{lupton02,scranton}.  In brief,
sources are classified as resolved or unresolved according to a measure of
light concentration that represents how well the flux distribution matches that
of a point source \citep{edr}.  As sources that emit strongly in the radio are
almost 
exclusively extra-galactic at the high latitudes observed by the SDSS
($|b|>30$), this classification effectively divides radio 
sources into ``galaxies'' (resolved) and ``quasars'' (unresolved).  The
optically-unresolved sources may also include a small number of galaxies that
are unresolved in the SDSS images, as well as a small fraction of radio stars.  
For the remainder of this paper, we refer to the two sets of photometric
optical sources (as opposed to those with available spectra, see below) as
``galaxies'' and ``optical point sources''. 

\subsubsection{The SDSS Spectroscopic Survey}
\label{subsubsec:sdss spectra}

A subset of photometric sources are chosen for spectroscopic observation
according to the SDSS's spectral target selection algorithms.  Targeted
extragalactic sources include the flux-limited ``main'' galaxy sample 
\citep[$r<17.77$;][]{strauss}, the luminous red galaxy sample
\citep{eisenstein}, and quasars \citep{richards02}.  DR6 contains spectra
for about 100,000 quasars and 790,000 galaxies.  The spectral wavelength range
is $3800-9200$\AA, with a resolution of 1800 and redshift accuracy of
30\,km\,s$^{-1}$ (estimated from the main galaxy sample).

The main galaxy sample includes nearly all ($\sim99\%$) galaxies brighter than
$r<17.77$, resulting in a sky density of $\approx90\deg^{-2}$.  Some targets
are rejected on the basis of low surface brightness (where redshifts become
unreliable or targets are spurious) or high flux (which can contaminate
neighboring fibers).  Due to the physical thickness of the spectral fibers, two
galaxies closer than $55\arcsec$ cannot be observed at the same time,
although overlapping spectral plates allow some initially-skipped galaxies to
be picked up later.  A second galaxy target algorithm selects luminous red
galaxies \citep[LRGs;][]{eisenstein}, which are typically the brightest members
of galaxy clusters.  LRG targets are selected primarily by color, based on
known LRG spectra at different redshifts.  The resulting sample is typically
more  luminous and much redder than the main galaxy sample, and extends to
fainter apparent magnitudes than the main sample. 

The quasar target selection algorithm \citep{richards02} selects targets from
unresolved objects with $i<19.1$ and colors similar to redshift $\lesssim3$
quasars, unresolved objects with $i<20.2$ and colors similar to higher redshift
quasars, and unresolved sources within $2\arcsec$ of a FIRST source ($i<19.1$).
The completeness of the sample is $\approx95\%$ and the selection efficiency is
$\approx66\%$.  Contamination is much higher than for the galaxy sample: the
latter consists of resolved sources, which are thus nearly always galaxies,
whereas the quasar sample is contaminated by non-quasar point sources, such as
distant galaxies or Galactic stars with quasar-like colors.  In fact, nearly
30\% of the targets turn out to be stars.

To distinguish the spectroscopic SDSS sample from the photometric SDSS sample,
we refer to sources from the former as ``spectroscopic galaxies'' and
``spectroscopic quasars''.

\section{THE UNIFIED CATALOG}
\label{sec:matching}

The cross-identification of radio surveys at different wavelengths, and with
different resolutions, is not a straightforward task
\citep{first,i02,devriesW,lu}.  However, the accurate astrometry of
FIRST allows simple positional matching to the other radio surveys, and
to the optical SDSS, with high completeness and low contamination.  Catalog
entries are defined by a detection in at least one of the two 20 cm surveys
(FIRST and NVSS). Where available, the 20\,cm data is supplemented with 92\,cm
and 6\,cm radio data, from WENSS and GB6, and with optical observations from
the SDSS.  

In the remainder of this section, we describe the matching technique
used to create the unified radio catalog.  The complicated procedure outlined
herein is intended to include all objects with even a small chance of
being physically-real associations.  Therefore, catalog users are not
limited to the matching techniques or matching distances used by the authors
in the subsequent analysis in this paper.  The generous matching radii used to
create the catalog can easily be restricted by future users using the catalog
parameters detailed in Appendix~\ref{app:matching}.

\subsection{Defining catalog entries from the FIRST and NVSS surveys}

FIRST and NVSS both observed the sky at 20\,cm (1400\,MHz).  Of the four radio
surveys incorporated into the unified catalog, these two have the faintest flux
limits and the highest spatial resolution, with FIRST having even 
higher resolution than NVSS.  Radio surveys necessarily involve a trade-off
between high resolution, which allows for accurate determination of positions,
and low resolution, which allows for the detection of low surface brightness
sources and complete flux measurements for extended sources.  By combining
high-resolution FIRST with the lower-resolution NVSS, it is possible to achieve
the best of both options.  FIRST provides accurate positional measurements, and
the NVSS provides accurate flux measurements for extended and low-surface
brightness sources, where FIRST underestimates the radio flux \citep{first,lu}.

\subsection{Matching algorithm for the radio surveys}
\label{subsec:match radio}

Due to different spatial resolution and faint limits, combined with the complex
morphology of radio sources, the positional matching of FIRST and NVSS sources
cannot be done in an a priori correct way.  For this reason, we adopt a
flexible method that enables later sample refinement after the initial
positional association.  The unified catalog contains the three closest FIRST
matches to an NVSS source within $30\arcsec$, and the three closest NVSS
matches to a FIRST source within $30\arcsec$.  For each source, we also record
the \emph{total number} of matches found within $5\arcsec$, $10\arcsec$,
$30\arcsec$, and $120\arcsec$, which will allow users to investigate radio
properties as a function of environment density.  The complete catalog contains
over 2.7\,million entries, including FIRST-NVSS associations,
isolated\footnote{\emph{Isolated} refers here to sources
  with no neighbors from the other 20\,cm survey within $30\arcsec$.} FIRST or
NVSS sources, and NVSS sources that lie outside the FIRST survey coverage.  
Because we match FIRST to NVSS and then NVSS to FIRST, there are
necessarily duplicate catalog entries: for example, a very close match will
appear once as a FIRST to NVSS match and once as a NVSS to FIRST
match. However, the catalog includes parameters to easily extract specific data
samples, including the elimination of any duplicates.  Similarly, the
processing flags can be used to treat cases such as distinct FIRST sources
matched to the same NVSS source (and vice-versa, although the latter case is
rare due to the lower spatial resolution and brighter limit of NVSS). For
further details, we refer the reader to Appendix~\ref{app:matching}.

\begin{figure}
\figcaption{\label{fig:first-nvss astrometry}
Differences in equatorial coordinate positions (\emph{top:} right
ascension; \emph{bottom:} declination) for FIRST-SDSS and NVSS-SDSS matches.
The full-width at half-maximum (FWHM) of the FIRST-SDSS distribution is about
$1\arcsec$.  For the NVSS-SDSS distribution the FWHM is about
$6\arcsec$.} 
\epsscale{0.7}
\plotone{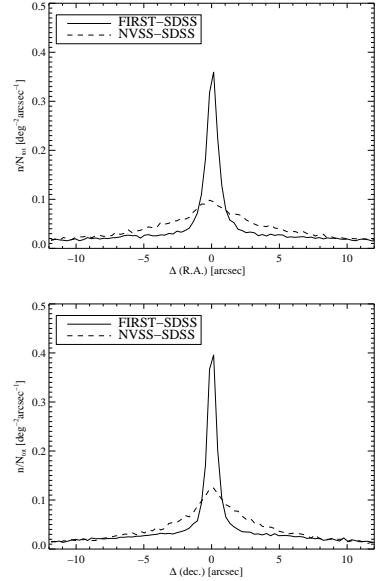}
\end{figure}

Measured FIRST and NVSS sky positions are rarely exact even for the same
source.  As shown in Figure~\ref{fig:first-nvss astrometry}, FIRST has more
accurate astrometric measurements than NVSS, due to its higher spatial
resolution.  We therefore use FIRST to designate the sky position of each
catalog entry, when possible.  For NVSS sources without a FIRST match, we retain
the NVSS coordinates.  The designated position is then used when searching for
counterparts in the other surveys.

To match the FIRST and NVSS surveys, we use a matching radius of 30\arcsec.  
We use a larger radius of $120\arcsec$ to match to WENSS and GB6, since they
both have lower positional accuracies.  We also record the total number of
WENSS and GB6 matches found within $120\arcsec$.  All of the matching radii
used to create the catalog are generously large, to ensure that the majority of 
physically-real matches are included.  To select smaller, cleaner samples for
analysis, the catalog user can choose matches based on distance, effectively
applying a smaller matching radius. The contamination and completeness as a
function of matching radius are discussed in \S\ref{subsec:completeness}. 

The surveys used in the creation of the unified catalog are themselves catalogs
of individual radio \emph{components}.  Thus, large multi-component physical
sources can be resolved into separate detections in the high angular-resolution
surveys, particularly in FIRST ($5\arcsec$ resolution), but also in NVSS and
WENSS ($\sim50\arcsec$ resolution).  \citet[][hereafter B05]{best} developed a
clean sample of double-lobed radio-loud galaxies by matching radio components
to spectroscopic SDSS galaxies.  B05 used the optical core position and the
optical galaxy alignment to eliminate unlikely lobe-configurations, based on
lobe opening angles and distances.  The complexity of a matching algorithm
based only on radio components, as presented in this paper, is necessarily much
greater because the optical position of the core is unknown for the majority of
sources.  We therefore settle on the matching scheme outlined above, which
allows sample selection with the user's own preferred matching criteria,
without the need to repeat the work that has gone into developing the unified
catalog presented here.

\subsection{Matching algorithm for the SDSS}
\label{subsec:match sdss}

We correlate the matched radio sources with the SDSS photometric survey,
including spectroscopic data when available, using a matching radius of
$60\arcsec$.  In addition to the nearest neighbor, the catalog includes the
brightest neighbor within a pre-defined radius ($3\arcsec$, $10\arcsec$,
$30\arcsec$, or $60\arcsec$; see Appendix~\ref{app:matching} for details).  The
total number of matches within the same pre-defined radius is also recorded,
indicating optical source sky density at each entry's location.

We note that some lobe-dominated radio sources may be missed when positionally
matching SDSS and FIRST source positions.  While each radio lobe will be
included individually in the radio sample, it will be excluded from the 
optically-matched subset for lobe-core distances larger than the SDSS matching 
radius.
\citet{lu} estimated that about 8.1\% of radio quasars do not show a
radio core within $2\arcsec$ of their optical position, and thus would be
missed by this algorithm.  However, \citet{i02} and \citet{devriesW} found
fewer than 5\% and 2\%, respectively, of quasars are double-lobed in FIRST,
using optical core positions without assuming radio emission in the core.
Because NVSS has a much larger spatial resolution than FIRST, even fewer 
double-lobed sources can be expected in NVSS.  B05 found 6\% of their
spectroscopic SDSS galaxies to be double-lobed in NVSS.  Note that while the
percentage is larger for galaxies, as expected from orientation effects, this
value is an upper limit for radio-optical samples using photometric optical
data, as the spectroscopic SDSS sample is comprised of the nearest sources,
which thus have large angular size.  Since the effect is not overwhelming, the
unified catalog only includes direct positional matches between SDSS and radio
(FIRST or NVSS) catalogs.  However, catalog users can easily repeat the
procedures for finding double-lobed sources developed by
\citet{i02,devriesW,best}, which would involve matching external optical
samples to the radio catalog sources.

\subsection{Completeness and efficiency of matched samples}
\label{subsec:completeness}

\begin{figure}
\figcaption{\label{fig:astrometry}
\emph{Diamonds} indicate distributions of distance between FIRST and (A) NVSS,
(B) WENSS, (C) GB6, and (D) SDSS positions for close pairs from sub-regions of
the catalog.  \emph{Squares} show the random match distribution, estimated by
off-setting the NVSS, WENSS, GB6, or SDSS positions by $1\degr$ in right
ascension.  Panel insets show the completeness (solid curve) and efficiency
(dotted curve) as a function of matching radius, estimated by a model fit to the
nearest neighbor distribution.  The solid vertical lines indicate radii used to
define catalog matches.  Dashed vertical lines show the more conservative
matching radii used for the analysis presented in this paper.}
\epsscale{0.8}
\plotone{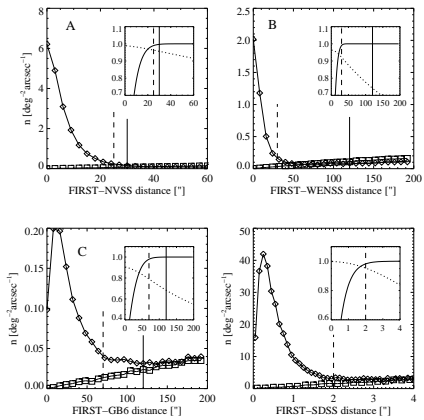}
\end{figure}

A side effect of using large matching radii is increased contamination by
coincidental ``line-of-sight'' matches to physically unrelated objects.
Using the nearest-neighbor distributions, we estimate efficiency (fraction of
matches which are physically real) and completeness (fraction of real matches
that were found) as a function of matching radius.  Figure~\ref{fig:astrometry}
shows the distribution of distances between FIRST sources and the nearest
neighbor from the NVSS, WENSS, GB6, and SDSS (photometric) surveys.  For
comparison, it also shows the estimated level of background contamination.  The
nearest-neighbor distributions peak at small distances, where associations are
likely to be real, then decrease sharply; the width of the peaks depends on the
astrometric accuracy of the corresponding surveys.  The distributions slowly
rise again at large distances due to increased background contamination.
Efficiency and completeness are estimated using a model fit to the nearest
neighbor distributions.  Estimated values of completeness and
efficiency based on the model fitting are listed in Table~\ref{table:matching
  radii}.  For the samples discussed here, typical values are $>90\%$
completeness and $>80\%$ efficiency.

As discussed above, the unified catalog includes individual radio components,
and does not correctly handle NVSS multi-component sources.  Two lobes from a
very large ($\gtrsim100\arcsec$) source may appear as two detections in NVSS
and thus in the unified catalog as well.  The values in
Table~\ref{table:matching radii} show that when matching up individual radio
components between different surveys, the completeness is well over 90\%.
As stated in \S\ref{subsec:match sdss}, investigations using optical and
radio surveys suggest that only a few per cent of radio sources are
double-lobed in FIRST; much smaller percentages are expected in the NVSS.
Indeed, B05 note that the NVSS resolution is large enough that $\sim99\%$ of
radio sources are contained in a single NVSS component.  Although the missing
double-lobed sources are not addressed further in this paper, we tackle the
issue in the companion paper, where we thoroughly discuss the classification of a
WENSS-NVSS-FIRST subsample (see \S\ref{subsubsec:wenss sample} for discussion).
It will include the missing double-lobed sources, found by determining the
complete NVSS-FIRST environment around each WENSS object, requiring extensive
visual analysis. 

\subsection{FIRST matching statistics}
\label{subsec:matching statistics}

Of the four radio surveys used to create the unified catalog, FIRST extends to
the faintest flux limit.  We discuss here the likelihood of finding a FIRST
source in one or more of the other three radio surveys.  In this section and
for the remainder of this paper, we limit our analysis to the common region
observed by all of the contributing surveys (see Fig.~\ref{fig:position
  plots}). 

The number of matches found between different surveys is a strong function
of source flux.  The flux range of objects in the unified catalog covers
several orders of magnitude.  Thus, for convenience we convert radio flux to an
``AB radio magnitude'', following \citet{i02}:
\begin{equation} 
\label{eq:magnitudes}
t_\mathrm{radio}=-2.5\log\left(\frac{f_\mathrm{int}}{3631\,\mathrm{Jy}}\right),
\end{equation} 
where $f_\mathrm{int}$ is the integrated flux density. This formula places the
radio magnitudes on the $\mathrm{AB}_\nu$ system of \citet{oke}.  An advantage
of that system is that the zero point (3631\,Jy) does not depend on wavelength
and thus enables convenient data comparison over a large wavelength range.
For example, a source with a radio flux of 1\,mJy has $t_\mathrm{radio}=16.4$; 
if it has constant $F_\nu$ (a flat spectrum), its visual magnitude is also
$V=16.4$.

\begin{figure}
\figcaption{\label{fig:matching statistics}
Sky density of FIRST sources matched to the other radio surveys as a function
of magnitude.  The total sky density of each
sample is indicated below the legend in each panel.  This plot includes FIRST
components from the same complex source; i.e. multiple FIRST detections matched
to a single NVSS source.
\emph{Panel A:} FIRST sources that were or were not detected in NVSS.  Sky
densities for sources brighter than 10\,mJy (13.9\,mag) are $18\deg^{-2}$
(\emph{triangles}) and $0.53\deg^{-2}$ (\emph{squares}). 
\emph{Panel B:} FIRST-NVSS sources with and without a counterpart in GB6.  Sky 
densities for sources brighter than 100\,mJy (11.4\,mag) are $1.5\deg^{-2}$
(\emph{triangles}) and $0.43\deg^{-2}$ (\emph{squares}).
\emph{Panel C:} FIRST-NVSS sources with and without a counterpart in WENSS.
Sky densities for sources brighter than 10\,mJy (13.9\,mag) are $16\deg^{-2}$ 
(\emph{triangles}) and $1.7\deg^{-2}$ (\emph{squares}).}
\epsscale{0.4}
\plotone{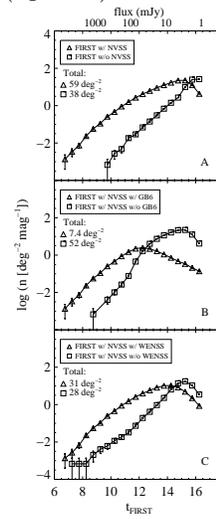}
\end{figure}

Figure~\ref{fig:matching statistics} shows the magnitude distribution for FIRST 
sources with and without radio matches from the other surveys.  Each panel
roughly indicates the $t_\mathrm{FIRST}$ magnitude corresponding to the other
survey's faint limit.  For example, the fraction of FIRST sources with a
counterpart in NVSS is greater than 0.9 at $t_\mathrm{FIRST}=15$ 
but drops to 0.5 at $t_\mathrm{FIRST}=15.5$ as shown in Panel~A, indicating
that the NVSS faint limit corresponds to $t_\mathrm{FIRST}\approx15$.  The 
sensitivity limit of GB6, $t_\mathrm{GB6}=13.3$, occurs at
$t_\mathrm{FIRST}\approx 12$ (panel B).  The limit is brighter in FIRST than
GB6 because, as is well known and as we confirm with a much larger sample in
\S\ref{subsec:colormag and colorcolor}, most radio sources are intrinsically
fainter at 6\,cm than at 20\,cm (in terms of $F_\nu$; i.e. corresponding
faint limits depend on spectral slope).  Requiring a GB6 detection therefore 
biases a sample to the brightest sources.  Panel~C shows that the
WENSS faint limit, $t_\mathrm{WENSS}=13.3$, corresponds to a limit of
$t_\mathrm{FIRST}\approx14$, because radio sources are, again on average,
intrinsically brighter at 92\,cm than at 20\,cm.

Roughly 60\% of FIRST sources have an NVSS counterpart within $30\arcsec$.
Of those radio sources observed by both surveys ($\approx48\deg^{-2}$),
50\% have a WENSS counterpart and 12\% have a GB6 counterpart within
$120\arcsec$, while 11\% have both.  The sky density of sources observed by
all four radio surveys is $\approx5.3\deg^{-2}$, or $\sim5\%$ of the FIRST
source density. 

\subsubsection{NVSS sources without a FIRST counterpart}

\begin{figure}
\figcaption{\label{fig:nvss astrometry}
\emph{Top:} Correlation between NVSS (lacking FIRST counterpart within
$30\arcsec$) and WENSS sources (\emph{diamonds}); random match distribution
(\emph{squares}) is shown for comparison, measured by offsetting WENSS
positions by $1\degr$ in right ascension.  The inset shows completeness
(\emph{solid curve}) and efficiency (\emph{dotted curve}) as a function of
matching radius.  Solid vertical lines indicate matching radius used to create
the catalog; dashed vertical lines show the matching radius used for this
paper's analysis. 
\emph{Bottom:} Sky density of NVSS sources without FIRST counterpart as a
function of magnitude.  The total sky density is indicated below the legend.
Sky densities for sources brighter than 10\,mJy (13.9\,mag) are 
$0.35\deg^{-2}$ (\emph{triangles}) and $0.21\deg^{-2}$ (\emph{squares}).}
\epsscale{0.5}
\plotone{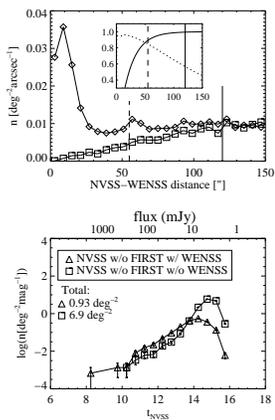}
\end{figure}

Because FIRST goes fainter than the NVSS by a factor of 2.5, there are many
FIRST sources without an NVSS counterpart ($\sim38\deg^{-2}$, see Panel~A of 
Fig.~\ref{fig:matching statistics}; $\sim0.53\deg^{-2}$ for FIRST sources
brighter than 10\,mJy).  However, the catalog also contains some NVSS sources
without a FIRST counterpart ($\sim7.8\deg^{-2}$; $\sim0.56\deg^{-2}$ for
sources brighter than 10\,mJy).  For an astronomical object to be detectable by
NVSS but not FIRST, it must be large and have a surface brightness too faint
for the high-resolution FIRST survey.  On the other hand, it is possible that
NVSS sources without FIRST counterparts are simply spurious, or have badly
measured NVSS positions, i.e. their FIRST counterparts are outside the
$30\arcsec$ matching radius.  The top panel of Figure~\ref{fig:nvss astrometry}
shows the nearest neighbor distribution from the WENSS catalog for NVSS sources 
\emph{lacking} a FIRST counterpart within $30\arcsec$.  It peaks at small
distances due to 
physically-associated objects, demonstrating that NVSS sources lacking a FIRST
counterpart but with a WENSS match (within $55\arcsec$), are dominated by real
sources.  If these are not spurious detections, they belong to the
category of extended, low-surface brightness objects missed by the
high-resolution FIRST. 

Assuming that real NVSS sources have a counterpart in at least one of FIRST and
WENSS, we estimate an upper limit for the fraction of spurious NVSS 
detections.  Nearly 14\% of NVSS objects in the region of survey overlap have
no FIRST counterpart within $30\arcsec$, and 12.2\% have no FIRST or WENSS
counterpart, the latter within $55\arcsec$.  NVSS has a much fainter flux
limit than WENSS (2.5\,mJy as opposed to 18\,mJy): 
sources near the NVSS faint limit are likely to fall below the WENSS faint
limit.  For the sample of NVSS detections brighter than 18\,mJy, only 0.43\%
lack both FIRST and WENSS counterparts, which we interpret as an upper limit
on the fraction of spurious NVSS sources.  The selection on which we base this
estimate is biased against sources brighter at 20\,cm than at 92\,cm; however,
this type of radio source is rare, as we later show (\S\ref{subsec:colormag and
  colorcolor}).  According to \citet{devries02}, NVSS begins to lose
completeness below 12\,mJy.  We find that 2.3\% of NVSS sources above this
limit have no FIRST match.

\subsection{Catalog availability}

The full catalog and several pre-made subsets are
available for download on the catalog
website\footnote{\emph{http://www.astro.washington.edu/akimball/radiocat/}},
which also provides a list and description of the data parameters.
The datasets are described in Appendix~\ref{app:data samples}.
All 2,724,343 rows of the complete catalog (sample~A) are available in a tarred
archive of \emph{fits} files, each covering a $5\degr$ wide strip in right
ascension.  A small subset of the catalog selected from $\approx100\deg^2$ of
sky, 
with 16,453 rows (sample~B), allows users to familiarize themselves with the
data format.  Several scientifically-useful subsets of the catalog are
pre-made for convenience; some of these are analyzed in the next section.  The 
first of these subsets contains radio fluxes and positions for 
NVSS-FIRST associations (sample~C), including 6 and 92\,cm data, while another
contains sources matched by all four radio surveys as well as the SDSS
(sample~F).  Two more subsets consist of spectroscopic galaxies (sample~G) and 
spectroscopic quasars (sample~H), detected by NVSS, FIRST, WENSS, and SDSS.  
Also provided are a sample of isolated\footnote{\emph{Isolated} here refers to
  an NVSS source with a single FIRST counterpart within $30\arcsec$.}
FIRST-NVSS sources (sample~I) and isolated FIRST-NVSS-SDSS sources (sample~J).
Finally, a set of high-redshift galaxy candidates is available (sample~K;
see~\S\ref{subsec:highz gals}).  A more detailed description of 
these data files can be found in Appendix~\ref{app:data samples}.

\section{                PRELIMINARY CATALOG ANALYSIS               }
\label{sec:analysis}

In this section, we investigate the distribution of sources in radio
color-magnitude-morphology parameter space.  We define the relevant
morphology and spectral parameters using five radio fluxes: peak FIRST flux
(20\,cm), integrated FIRST and NVSS fluxes (20\,cm), GB6 flux (6\,cm), and
WENSS flux (92\,cm). While the unified catalog contains many more useful
parameters inherited from the original catalogs, in this preliminary analysis
we focus only on these few.  We use FIRST and NVSS flux measurements
to define two morphology estimators, and use GB6 and WENSS fluxes
with NVSS flux to compute radio spectral slopes.  Morphology
estimators are used to define three  morphology classes, for which we
construct radio ``color-magnitude'' diagrams.  We also utilize optical
photometric identification in the analysis, classifying sources as SDSS point 
sources, extended sources, or faint sources (non-detections).

Hereafter our discussion covers samples selected using the conservative
matching radii listed in Table~\ref{table:matching radii}.  We settle upon 
these values by looking for an optimal balance between completeness
and efficiency of the matched samples, as illustrated in
Fig.~\ref{fig:astrometry}.  The matching radii used herein are as follows, with
the larger values for all catalog associations given in 
parentheses: FIRST-NVSS, $25\arcsec$ ($30\arcsec$); FIRST-WENSS, $30\arcsec$
($120\arcsec$); FIRST-GB6, $70\arcsec$ ($120\arcsec$); FIRST-SDSS, $2\arcsec$
($60\arcsec$).  Estimated values of completeness and efficiency, calculated by
comparing with random matching, are listed in Table~\ref{table:matching radii}.

Our analysis is limited to
the $2955\deg^2$ region where the contributing surveys overlap
(Fig.~\ref{fig:position plots}) in
order to control the selection criteria of the analyzed samples.
There are nearly 490,000 catalog entries in the overlap region, including
multiple FIRST components matched to the same NVSS source, as well as duplicate 
entries (see~\S\ref{subsec:match radio}).  To obtain a sample of
physical FIRST-NVSS associations, it is sufficient to select catalog entries
containing an NVSS source and its nearest FIRST match: individual objects are
typically not resolved into multiple components in NVSS because of the survey's
larger beam.  This method selects $\sim140,000$
unique, likely physical NVSS-FIRST associations.  Of those, $\sim64,000$ are
matched to a WENSS source, $\sim14,000$ to a GB6 source and $\sim48,000$ to an
SDSS source.

\subsection{Morphology classification of radio sources}
\label{subsec:morphology}

\subsubsection{     ``Simple'' vs. ``complex'' FIRST-NVSS sources }

\begin{figure}
\figcaption{\label{fig:I02 fig13}
\emph{Top:} Difference in 20\,cm radio magnitudes measured by FIRST and NVSS as
a function of $t_\mathrm{NVSS}$.  Where individual points saturate the plot, we 
show the distribution using contours.  The dashed horizontal lines indicate
peaks at $\Delta t=0$ and $\Delta t=0.7$.  The solid horizontal line shows the
$\Delta t=0.35$ separator between between complex and simple sources.
Vertical dashed lines mark the magnitude bins shown in the lower panel.
\emph{Bottom:} Distribution in $\Delta t$ for four $t_\mathrm{NVSS}$ bins.  The
vertical solid line shows the $\Delta t=0.35$ separator between complex and
simple sources.}
\epsscale{0.5}
\plotone{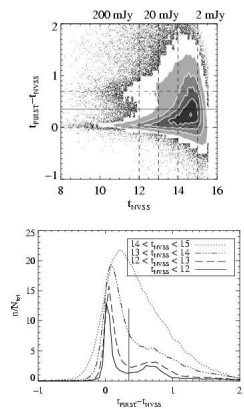}
\end{figure}

FIRST is a high-resolution interferometric survey, and thus underestimates the
flux of extended and lobe-dominated sources \citep{first,lu}.  Additionally, a
multiple-component source with core and lobes may be detected as three objects
by FIRST but as only a single object in the lower-resolution NVSS.  The
difference between the two 20\,cm magnitudes is thus a measurement of source
morphology that indicates angular extent and complexity.  We define 
\begin{equation}
\label{eq:delta t}
\Delta t = t_\mathrm{FIRST} - t_\mathrm{NVSS}.
\end{equation}
Before applying $\Delta t$ as a classifier we investigate its dependence on
$t_\mathrm{NVSS}$ for NVSS-FIRST pairs (sample~C) in
Figure~\ref{fig:I02 fig13}.  The distribution in $\Delta t$ is bimodal, with
the two peaks centered at $\Delta t=0$ and $\Delta t=0.7$.  Through
the visual inspection of a thousand FIRST images (see below,
\S\ref{subsubsec:auto class}), we verified that sources in the $\Delta 
t\sim$ locus are single component sources, while those in the $\Delta
t\sim.7$ locus are 
multiple-component or extended.  The lower panel shows the $\Delta t$
distribution for four magnitude bins.  We fit the sum of two Gaussians to each
of the four distributions; best-fit parameters are listed in
Table~\ref{table:gaussfits}.  The $\Delta t$ distribution is similar for all
magnitudes, although the overlap between the two peaks increases for fainter
sources as flux measurement errors increase.  We adopt $\Delta t = 0.35$ (solid
horizontal line) as the separator between ``simple'' ($\Delta t <0.35$) and
``complex'' ($\Delta t>0.35$) sources.  

\subsubsection{     ``Unresolved'' vs. ``resolved '' FIRST sources }

The ratio of the two FIRST flux measurements---\emph{peak} and
\emph{integrated}---yields a second measure of morphology: a
dimensionless source concentration on $\sim5\arcsec$ scale.  We define
\begin{equation}
\label{eq:first size}
\theta=\left(\frac{F_\mathrm{int}}{F_\mathrm{peak}}\right)^{1/2}. 
\end{equation}
Sources with $\theta\sim$ are highly concentrated (unresolved), while
sources with larger $\theta$ are extended (resolved).  We adopt
$\log(\theta^2)=0.05$ ($\theta\approx1.06$)
as the value separating resolved and unresolved sources.  This choice is
motivated by the distribution of sources in the two-dimensional $\Delta t$ vs. 
$\theta$ distribution, discussed below.

\subsubsection{  Automatic morphology classification of radio sources }
\label{subsubsec:auto class}

\begin{figure}
\figcaption{\label{fig:delt v size}
\emph{Top:} Distribution of $\Delta t$ for sources with $t_\mathrm{NVSS}<13.5$.
The dashed line indicates the separation between the \emph{simple} and
\emph{complex} morphology classes. 
\emph{Middle:} Distribution of $\theta$, an estimate of source size.  The solid
curve corresponds to simple sources; the dotted curve corresponds to
complex sources.  The vertical line separates \emph{unresolved} and
\emph{resolved} sources at 20\,cm.
\emph{Bottom:} Distribution of $\Delta t$ vs. $\log\theta^2$ for sources with
$t_\mathrm{NVSS}<13.5$.  The dashed lines indicate the adopted separation of 
morphology classes: the complex sample is above the horizontal dashed line,
the compact sample is left of the vertical dashed line, and the
resolved sample is to the right of the vertical dashed line.}
\epsscale{0.5}
\plotone{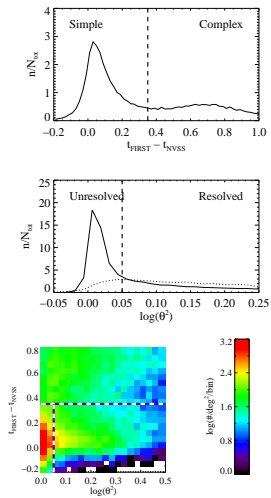}
\end{figure}

The two-dimensional $\Delta t$ vs. $\theta$ distribution for sources with
$t_\mathrm{NVSS}<13.5$ (the brightest 30\% of NVSS-FIRST pairs) is illustrated in
Figure~\ref{fig:delt v size}, along with the marginal distributions for the two
morphology parameters.  The distribution in the lower panel
suggests that 20\,cm radio fluxes can be used to automatically separate
FIRST-NVSS detections into three morphology
classes: ``complex'', (simple) ``resolved'', and (simple) unresolved or
``compact''. 

\begin{figure}
\figcaption{\label{fig:sources}
Examples of radio sources classified by morphology (FIRST
$1\arcmin\times1\arcmin$ stamps normalized to the brightest pixel, shown with
square root stretch).  The top row contains \emph{compact} radio sources, the
second row shows \emph{resolved} sources, and the remainder are \emph{complex}.
The left two columns show sources optically identified as galaxies by the SDSS;
the right two columns show SDSS-identified quasars. }
\plotone{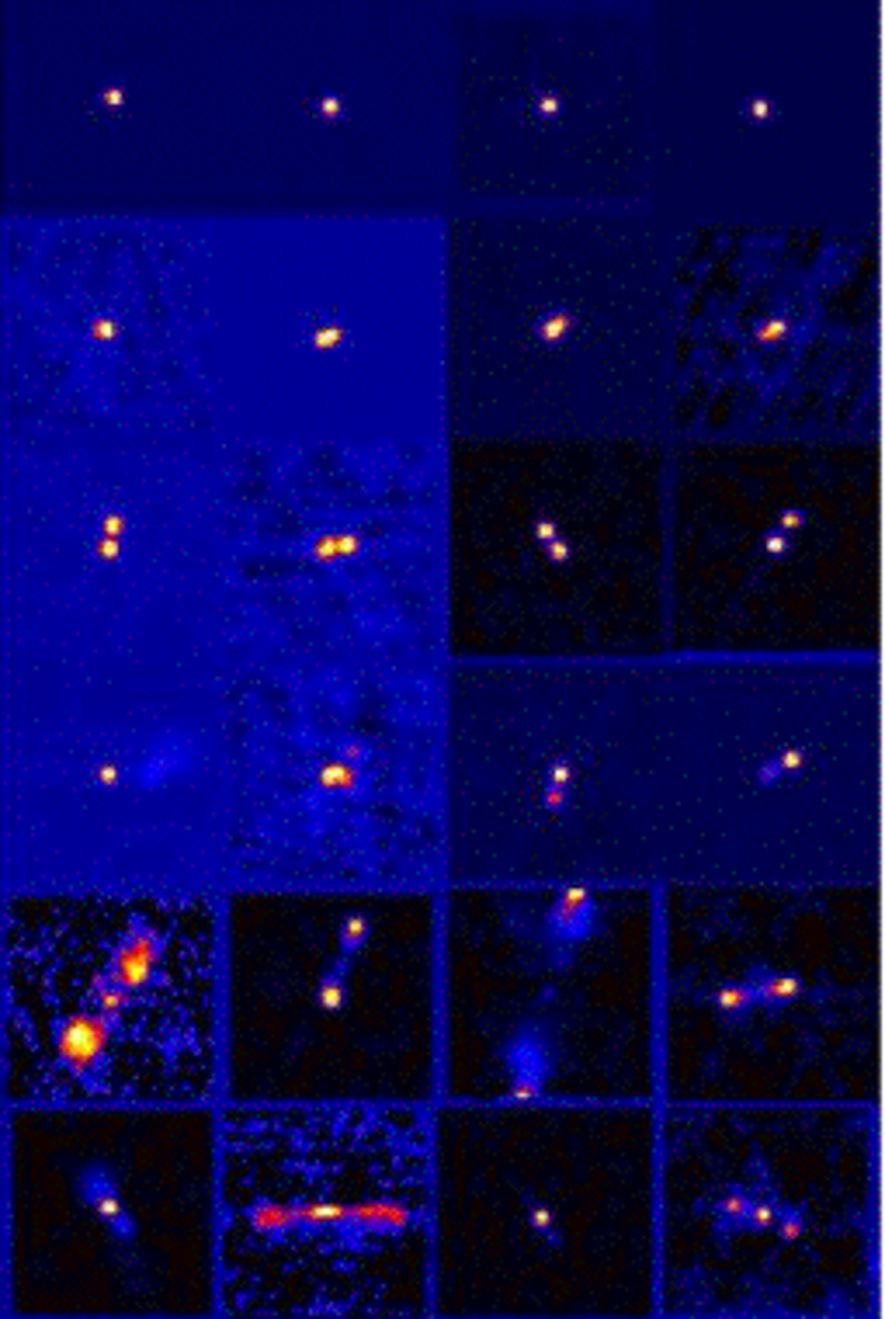}
\end{figure}

This assertion was verified by extensive visual inspection of FIRST images
(Fig.~\ref{fig:sources}).  
Over 1000 FIRST stamps ($2\times2~\mathrm{arcmin}^2$) of optically-identified
radio quasars and radio galaxies were classified both visually and
automatically (as above) into ``complex'', ``resolved'', and
``compact'' categories.  A comparison of the results is given in
Table~\ref{table:morphology}.  The two methods are consistent:
over three quarters of all sources (81\% of quasars and 76\% of galaxies)
receive identical classification from the two methods.  Two initially surprising
results are the significant fraction of objects classified as
``complex'' by one method and ``compact'' by the other.  In particular, 15\% of
visually-complex quasars were automatically classified as ``compact''.  As it
turns out, the majority of these sources appear (by eye) to be asymmetric
double-lobed sources with a very high flux ratio.  Their FIRST images show two 
lobes, hence the visual ``complex'' classification.  However, their total flux
is due primarily to the brighter lobe, meaning FIRST and NVSS flux measurements
are similar, resulting in an automatic classification of ``compact''.
Additionally, 26\% of visually-compact galaxies were classified automatically
as ``complex''.  An inspection of flux values reveals that many of these are
borderline cases, with $\Delta t$ value just above the $\Delta t=0.35$
cutoff.  It is likely that these galaxies have faint, diffuse emission which is
detected by the NVSS but is not visible in the FIRST images.

The automatic morphology classification presents difficulties because of
non-unique pair matching, e.g. multiple FIRST detections matched to a single
NVSS object.  The resolved/unresolved classification is based on FIRST
fluxes and therefore applies to FIRST components individually.  In choosing an
analysis sample based on FIRST-NVSS pair matching, we retain only the closest
FIRST match to an NVSS object (see Appendix~\ref{app:matching} for details).
The fluxes of the closest FIRST match are used, while more distant FIRST
sources are ignored, when classifying a FIRST-NVSS pair as resolved/unresolved.
Approximately 17\% of NVSS detections in the FIRST footprint have multiple
FIRST matches within 30\arcsec.  Most of these sources (88\%) are classified as
complex, presumably because the NVSS flux encompasses multiple FIRST
components.  We therefore apply the resolved/unresolved classification only to
``simple'' sources, resulting in the three morphology classifications.  (In
other words, we do not separate the complex sources into ``complex resolved''
and ``complex unresolved''.) 

\begin{figure}
\figcaption{\label{fig:logN morphology}
\emph{Top:} Distribution of NVSS-FIRST sources classified according to radio
morphology, as a function of $t_\mathrm{NVSS}$. 
\emph{Bottom:} Fraction of sources brighter than $t_\mathrm{max}$ belonging
to each morphology class.}
\plotone{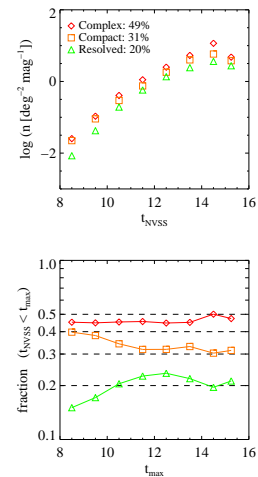}
\end{figure}

Figure~\ref{fig:logN morphology} shows the magnitude distribution for
NVSS-FIRST associations of each morphology class, and the fraction in each
class of all sources brighter than a limiting $t_\mathrm{NVSS}$ magnitude.
There are similar numbers of complex and compact sources at all magnitudes;
there are fewer resolved sources, and the fraction increases slightly at
fainter flux values.  Overall, the fractions remain roughly constant with
flux.  In the next section, we investigate correlations between radio
morphology and radio spectral slope.

\subsection{Radio color-magnitude and color-color diagrams}
\label{subsec:colormag and colorcolor}

In the previous section, we described an automatic morphology classification
for NVSS-FIRST associations based on their 20\,cm fluxes.  The addition of
WENSS (92\,cm) and GB6 (6\,cm) data allows us to compute radio spectral slope,
and to study its correlation with radio morphology and magnitude.

For two magnitudes $t_1$ and $t_2$, we compute a spectral index $\alpha$,
defined by $F_\nu\propto\nu^\alpha$, as 
\begin{equation}
\label{eq:alpha}
\alpha^{t_1}_{t_2}=\frac{0.4}{\log(\lambda_{t_1}/\lambda_{t_2})}(t_1-t_2).
\end{equation}
The value $\alpha$ describes the average spectral slope between wavelengths
$\lambda_{t_1}$ and $\lambda_{t_2}$; it is proportional to the color
$t_1-t_2$.  Therefore 
\begin{equation}
\alpha^{20}_6=0.765(t_\mathrm{NVSS}-t_\mathrm{GB6}),
\end{equation}
\begin{equation}
\alpha^{92}_6=0.337(t_\mathrm{WENSS}-t_\mathrm{GB6}),
\end{equation}
and
\begin{equation}
\label{eq:alpha9220}
\alpha^{92}_{20}=0.604(t_\mathrm{WENSS}-t_\mathrm{NVSS}).
\end{equation}
Note that we use NVSS rather than FIRST when
calculating spectral indices for the analysis presented here, as NVSS is
more likely to detect all the 20\,cm flux from a source due to its larger
beamsize.

Typical extragalactic radio sources have spectral index in the range
$-1<\alpha<0$ in the radio regime \citep[e.g.,][]{condon,brw}, with
$\alpha=-0.5$ as a typical value separating steep-spectrum and
flat-spectrum sources \citep[e.g.,][]{urry_review}.  Emission
from extended radio lobes, such as those seen in classical FR2 objects
\citep{FR} tends to be steeper, with the steepening attributed to synchrotron, 
inverse Compton, and adiabatic losses acting on the relativistic electrons.
The flatter emission from compact quasar cores/jets is interpreted as
self-absorbed synchrotron emission \citep{krolik}.  
As described in \citet{jarvis}, radio spectral index is thought to correlate
with source orientation in AGN unification schemes: sources viewed along the
jet axis show flat radio spectra, sources viewed along the dust torus show
steep radio spectra (since only the lobes are visible), and intermediate
orientations result in intermediate spectral slopes due to mixing of the flat-
and steep-spectrum emission.  A minority of observed radio sources fall
into the gigahertz-peaked-spectrum (GPS; 10\%) and compact-steep-spectrum (CSS;
30\%) categories \citep{o'dea}.  GPS and CSS sources are similar to smaller
versions of FR2s, but have convex spectra that peak around 1\,GHz in
the observer frame.\footnote{GPS galaxies are thought to be young, evolving
  radio sources that expand through a CSS stage and eventually turn into FR1 or
  FR2 galaxies \citep{o'dea,devriesN}.  GPS and CSS sources that peak at
  sufficiently short wavelengths can have positive $\alpha$ in the unified
  catalog.} 

Two significant systematic issues could cause errors in our measure of radio
spectral index.  The first of these is the possible
variability in radio flux over the timescale of the survey observations: GB6
observations were carried out in 1986 and 1987, WENSS operated in the early
1990s, and NVSS observations were taken in the first half of 1998.
AGN have shown flux variations over all radio wavelengths on timescales ranging
from less than a day to years.  Intrinsic variability is thought to be due to
jet variations, such as shocks propagating along the jet
\citep{marscher,hughes}, with Doppler boosting acting to exaggerate the effect.
Therefore, variability should be more prevalent in sources viewed along or near
the jet axis, which is consistent with observations:
blazars, quasars, flat-spectrum sources, and compact sources show higher
variability among a greater fraction of sources than control samples.  For
example, \citet{barvainis} observed variability at the 5-8\% level over a
two-year timescale for core-dominated quasars observed at 8.4\,GHz.  They also
observed some large variations on month-long timescales, but only one in five of
their sources varied at the 10-30\% level, and most varied by less than 20\%.
Meanwhile, \citet{devries04} found that roughly 2\% of their sample of
unresolved FIRST sources show significant variability ($>4\sigma$).  In a sample 
of 3600 sources brighter than 0.1\,mJy at 1.4\,GHz, \citet{rys} found that less
than 1\% were variable, and that \emph{all} of those have flat radio spectra,
with 86\% being core-dominated sources.
As far as measurements for individual sources are concerned, a 10\% flux
variation at one of the two relevant wavelengths corresponds to a change of
0.08 in spectral index, while a 30\% flux 
variation corresponds to a change of about 0.2 in spectral index.  Results
from the literature discussed above suggest that about 1\% of the unified
catalog sources varied significantly over the range of survey observations, and
that the variation tended to be at the 10\% level.  Those sources are likely to
be concentrated in the compact radio morphology sample with
flat radio spectra ($\alpha\gtrsim-0.5$).  This variability contributes to 
scatter in the distribution of $\alpha$.

The second issue which may cause erroneous calculations of spectral index is
the fact that the radio surveys have different spatial resolutions.  WENSS and
NVSS both have $\sim50\arcsec$ resolution (see
Table~\ref{table:surveys}), while GB6 has a resolution of about
$200\arcsec$, and therefore gathers flux from a much larger area.  
This issue will effect $\alpha^{20}_6$ spectral index for radio
sources resolved into separate NVSS components, such as a double-lobed radio
galaxy with lobe separation $\sim100\arcsec$.
If a single NVSS component contains only half of a source's total 20\,cm
flux while GB6 sees all of the 6\,cm flux, then $\alpha^{20}_6$ 
will be too shallow by almost 0.6 for that source.  We note that since WENSS
and NVSS have similar angular resolution, this issue is unlikely to
significantly affect the $\alpha^{92}_{20}$ measurements.

\begin{figure}
\figcaption{\label{fig:radio colorcolor}
Color-magnitude (\emph{left}), color-color (\emph{middle}), and median
magnitude (\emph{right}) diagrams for sources detected in all four radio
surveys and classified by radio
morphology (\emph{top:} complex; \emph{middle:} resolved; \emph{bottom:}
compact).  Vertical dashed lines show the $\alpha^{92}_{20}=-0.5$ separation
between steep- and flat-spectrum sources in the 20-92\,cm range.
\emph{Left column:} The distribution of NVSS magnitude vs. spectral slope
$\alpha^{92}_{20}$.
\emph{Center column:} $\alpha^{20}_6$ vs. $\alpha^{92}_{20}$ color-color
diagrams.  The diagonal, dashed red lines show the
$\alpha^{92}_{20}=\alpha^{20}_6$ locus. 
\emph{Right column:} Color-color diagrams with bins color-coded according to
the median value of NVSS magnitude $t_\mathrm{NVSS}$.  The strong gradients in
the right column are due to steep-spectrum sources, which are biased to bright
sources by the shallow GB6 survey.}
\epsscale{0.9}
\plotone{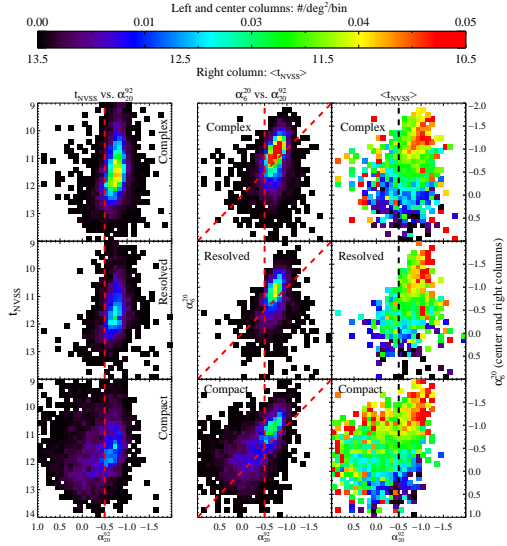}
\end{figure}

Figure~\ref{fig:radio colorcolor} presents color-magnitude and color-color
diagrams for the three radio morphology classes detected in all
four radio surveys (listed as sample~E in Appendix~\ref{app:data samples}).
Similar color-color diagrams to those in the 
middle column were first presented in \citet[][Fig.~2]{ivezic04}, with slightly
different selection criteria.  The middle column 
demonstrates that the majority of sources in all three morphology classes have
spectral indices $\alpha^{92}_{20}\approx\alpha^{20}_6\approx-0.8$.
This result is in agreement with earlier observations
\citep[e.g.,][]{kellermann,laing,zhang}, and suggests that most strong radio
sources have a fairly constant power-law slope 
from 6 to 92\,cm.  However, the radio compact sample (\emph{lower middle
  panel}) has a significant fraction of flat-spectrum
sources ($\alpha\sim0$) as well.  In the next
section, we investigate the two groups of compact sources using optical
identifications, and show that they are due to a population of galaxies and a
population of optical point sources, with different distributions in radio
color-color space.  Note that in the color-color diagrams, the WENSS faint
limit biases against large positive $\alpha$, while the
GB6 limit biases against large negative~$\alpha$.  The strong correlation
between $\alpha^{20}_6$ and $\left<t_\mathrm{NVSS}\right>$ in the right column
is due to the GB6 survey sensitivity: only very bright steep sources are
detected by GB6.  A small fraction of sources ($<2\%$ with
$t_\mathrm{NVSS}=12$) have a negative value for $\alpha^{92}_{20}$ but a
positive value for $\alpha^{20}_6$, i.e. are fainter at 20\,cm than at 92\,cm
and 6\,cm.  Such sources either have highly unusual radio spectra, or varied 
significantly in brightness over the range of observations.

\begin{figure}
\figcaption{\label{fig:logN slope}
The magnitude distribution of compact morphology radio sources, observed by all 
four radio surveys and brighter than $t_\mathrm{NVSS}=12$, categorized into
four subclasses by spectral shape.  
\emph{Top:} Representations of a standard spectral shape in each class, defined
by radio fluxes at 6, 20, and 92\,cm.  \emph{Steep} and \emph{flat} refer to
sources with monotonic radio flux measurements.
\emph{Steep} sources have $\alpha^{92}_{6}<-0.5$; \emph{flat} sources have
$\alpha^{92}_{6}>-0.5$.  \emph{Peaked} refers to sources which are brightest at
20\,cm.  \emph{Inverted} refers to sources which are faintest at 20\,cm.
\emph{Middle:} Twenty centimeter magnitude distribution for each spectral shape 
subclass.
\emph{Bottom:} Fraction of sources brighter than $t_\mathrm{max}$ belonging to
each spectral shape class.} 
\epsscale{0.44}
\plotone{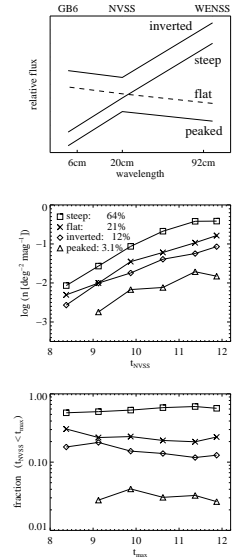}
\end{figure}

We quantify the fraction of compact sources with steep and flat spectra
in Figure~\ref{fig:logN slope}.  The sources are divided
into four subclasses depending on their spectral shape between 6\,cm and
92\,cm.  The top panel shows an example shape for each subclass.  The ``steep''
and ``flat'' designations refer to sources with monotonic radio flux
measurements.  The middle panel shows the magnitude distribution
of each subclass for compact sources with
$t_\mathrm{NVSS}\leq12$.  Slightly less than two thirds of the compact 
sources are steep-spectrum, whereas 94\% of complex and resolved sources are
steep-spectrum.  We find that one quarter of compact sources are flat-spectrum
or peaked between 6\,cm and 92\,cm; these values are consistent with the
fractions of GPS and CSS sources found by \citet{o'dea}.  The fractions of 
steep, flat, and peaked sources remain nearly constant with magnitude
(\emph{bottom panel}).  However, \citet{devries02} observed that the ratios
change at fainter fluxes, where the fraction of flat-spectrum sources rises
sharply (their Fig.~1).  Note that because of the flux limits of the
contributing radio surveys, we are more sensitive to inverted-spectrum sources
than to peaked-spectrum sources.

\subsection{Optical identification of radio sources}
\label{subsec:optical ids I}

The addition of SDSS data enables the separation of radio sources into galaxies 
(i.e., extended optical sources), optical point sources, and faint optical
sources (i.e. undetected by SDSS).  Combining the 
three optical categories with the three radio morphology categories
yields nine optical/radio classes.  As discussed in
\S\ref{subsubsec:sdss photometry}, the sample of optical point sources may
include a small number of unresolved galaxies and radio stars in addition to
radio quasars.

\subsubsection{The NVSS-FIRST-WENSS-GB6 sample}

We begin by discussing the approximately 12,000 sources detected by 
\emph{all four radio surveys} (required in order to measure the two radio
spectral indices $\alpha^{92}_{20}$ and $\alpha^{20}_6$).  Over 38\% of these
have an SDSS counterpart within $2\arcsec$.  This is slightly larger than the
fraction of FIRST sources with an optical counterpart (32\%), and may be due to
a bias towards the bright end induced by requiring GB6 and WENSS detections.
This dataset is listed as Sample\,E in Appendix~\ref{app:data samples}; the
optically-matched subset is Sample\,F. 

\begin{figure}
\figcaption{\label{fig:sdss matching statistics}
Magnitude distributions for sources detected by all four radio surveys.
\emph{Panel~A:} Full sample divided by radio morphology.  
\emph{Panel~B:} Full sample divided by optical identification.
\emph{Panel~C:} Optically-identified subset divided by radio morphology.
\emph{Panels~D-F:} Complex, resolved, and compact radio sources,
respectively.}
\epsscale{0.85}
\plotone{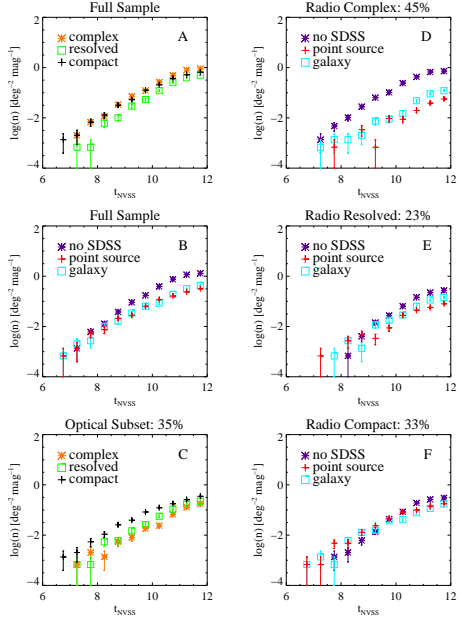}
\end{figure}

Figure~\ref{fig:sdss matching statistics} shows $t_\mathrm{NVSS}$ magnitude
distributions for sources detected by all four radio surveys and brighter than
$t_\mathrm{NVSS}=12$, corresponding roughly to the GB6 detection limit (see
\S\ref{subsec:matching statistics}).  For the optically-matched subset, most
sources have compact radio morphology, while the fraction of complex
sources drops dramatically.  This is partly due to the fact that
radio-optical matched samples are biased against double-lobed sources
\citep{lu}.  The fractions of sources in each morphology class remain roughly
constant with $t_\mathrm{NVSS}$ even for the optically-matched subset.  In all
three radio morphology classes, galaxies and point sources are found in roughly
equal ratios at all magnitudes.

\begin{figure}
\figcaption{\label{fig:radio colorcolor optical}
Radio color-color diagrams for sources detected by all four radio surveys
divided into the nine radio/optical classes.
The abcissa is
$\alpha^{92}_{20}$; the ordinate is $\alpha^{20}_6$.  From top to bottom, rows
contain complex, resolved, and compact radio sources.  From left to right the
columns contain optically undetected sources, optically resolved sources, and
optically unresolved sources.  Contours in each column show the underlying
distribution of all sources in the corresponding column.  Contour levels and
histogram normalization are arbitrary for each column.}
\epsscale{0.85}
\plotone{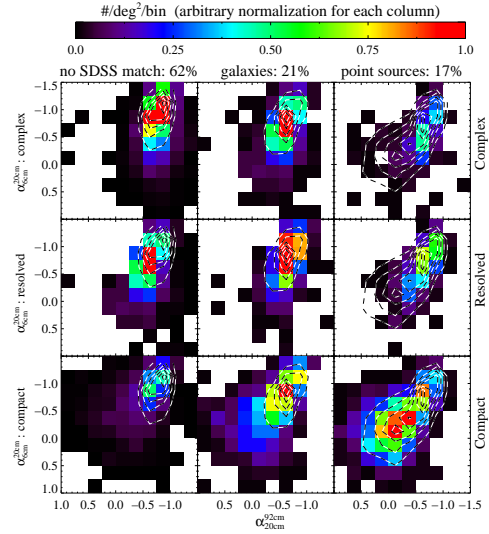}
\end{figure}

Figure~\ref{fig:radio colorcolor optical} shows radio color-color diagrams for
the nine radio morphology/optical identification classes.  This figure is an
expansion of the middle column of Figure~\ref{fig:radio colorcolor}, with the
addition of the three optical identification categories.  Faint sources,
galaxies, and point sources show different color-color distributions in the
three morphology classes, particularly in the compact sample.  For the complex
sample, faint optical sources peak at steep values of $\alpha\sim-1$
while point sources peak near $\alpha\sim-0.5$.  Galaxies are more
tightly clustered around the peak than point sources.  The
sharpest separation in  behavior is seen in the compact morphology sample
(\emph{bottom row}).  The optically-faint and galaxy samples tend to have steep
spectral indices ($\alpha<-0.5$), while point sources tend to have flat
spectral indices ($\alpha>-0.5$).  This confirms, \emph{with much improved
  statistics}, the long-known distinction between flat spectrum radio quasars
which are dominated by core emission and steep spectrum radio quasars which are
dominated by lobe emission \citep{krolik}.  The compact point sources are also
bimodal, with both a flat-spectrum and a steep-spectrum peak, while point
sources in the complex and resolved morphology classes are primarily
steep-spectrum.  The steep-spectrum peak in the compact morphology class
(\emph{lower right panel}) may be due to distant, unresolved galaxies
contaminating the point source sample.

\begin{figure}
\figcaption{\label{fig:alphamean all radio}
Distribution of the average of spectral indices $\alpha^{20}_6$ and
$\alpha^{92}_{20}$ for $t_\mathrm{NVSS}<12$ sources detected in
all four radio surveys.  The vertical dashed lines show the separation between 
steep- and flat-spectrum sources.  Note that the Y-axis range varies.
\emph{Panel~A:} Full sample divided by radio morphology.  
\emph{Panel~B:} Full sample divided by optical identification.
\emph{Panel~C:} Optically-identified sources divided by radio morphology.
\emph{Panels~D-F:} Complex, resolved, and compact radio sources,
respectively.}
\epsscale{0.77}
\plotone{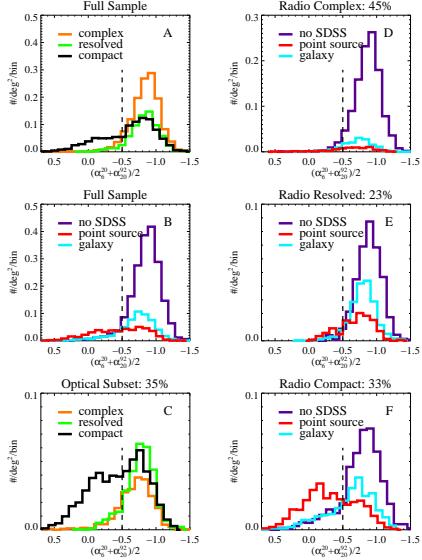}
\end{figure}

The separation of the steep- and flat-spectrum peaks in color-color space
is maximized along the $\alpha^{92}_{20}=\alpha^{20}_6$ locus.  We thus take
the average of the two indices to quantitatively investigate spectral slope
distributions, as presented in
Figure~\ref{fig:alphamean all radio}.  Complex and radio resolved classes
contain primarily steep spectrum sources, while the optically-identified radio
compact sample is bimodal with a peak due to flat-spectrum point sources and a
peak due to steep spectrum galaxies.  The optically-faint subset of the compact
morphology class is biased toward steep-spectrum sources (compare panels A
and~C).  \emph{The distributions for the optically faint sample are similar to
  those of the galaxy sample.}  These histograms are based on the largest
multi-wavelength samples of radio sources ever constructed.
Table~\ref{table:alphamean} compares the mean, median, and FWHM for the
distributions in panels A, D, E, and~F. 

\subsubsection{The NVSS-FIRST-WENSS sample}

The majority of radio sources have $\alpha<-0.5$, i.e. they are brighter at
longer radio wavelengths.  
The sample of sources detected by all four radio surveys is therefore flux 
limited by the short wavelength GB6 survey.  Dropping the requirement of a
detection in GB6 increases the number of selected sources to 64,000, a factor
of five, while spectral slope information is still available from
$\alpha^{92}_{20}$.  The $\alpha^{92}_{20}$ distributions for the larger sample
of NVSS-FIRST-WENSS detections (not shown) include much fainter sources; the
observed bimodality seen in Figure~\ref{fig:alphamean all radio} is still
present, although the separation is weaker.  Mean, median, and FWHM of the
$\alpha^{92}_{20}$ distributions for the samples in Figure~\ref{fig:alphamean
  all radio} are shown in Table~\ref{table:alpha9220 gb6}.  We use the
NVSS-FIRST-WENSS subsample (listed as sample~D in Appendix~\ref{app:data
  samples}) for all further data analysis.

\subsubsection{Summary of source characteristics in the unified radio catalog}

\begin{figure}
\figcaption{\label{fig:radio colormag}
Radio color-magnitude diagrams for the nine radio/optical subclasses from the
NVSS-FIRST-WENSS sample, categorized by radio morphology and optical
identification.  From top to bottom, the rows contain complex, resolved, and 
compact radio sources.  From left to right, the columns contain the 46,517 
optically undetected, the 12,780 optically resolved, and the 4,363 optically
unresolved sources.  The left axis is labeled with radio magnitude; the right
axis with corresponding flux.  Contours correspond to all the
sources in each column; contour levels and histogram normalization are
arbitrary for each column.  Note that radio sources without an SDSS match are
dominated by complex steep sources ($\alpha^{92}_{20}<-0.5$), while those
matched to an SDSS point source are dominated by compact flat sources
($\alpha^{92}_{20}>-0.5$).} 
\epsscale{0.8}
\plotone{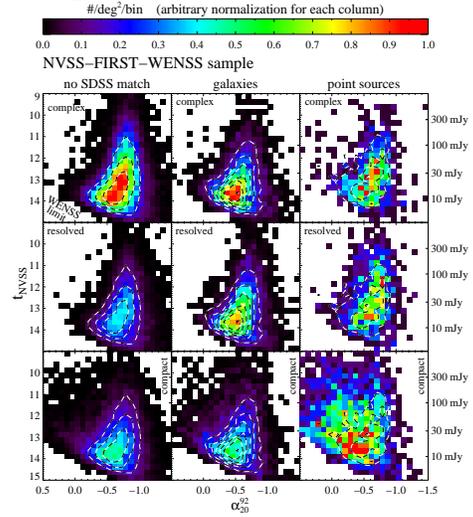}
\end{figure}

Figure~\ref{fig:radio colormag} shows radio color-magnitude diagrams for
NVSS-FIRST-WENSS sources from the catalog overlap region, providing a
qualitative summary of 
fundamental properties of radio sources in the unified catalog.  This figure
expands the left panel of Figure~\ref{fig:radio colorcolor} by adding classes
defined by the optical identification categories.  However, Figure~\ref{fig:radio 
  colorcolor} was limited to sources observed by GB6 in addition to NVSS,
FIRST, and WENSS, resulting in a sample of only the brightest radio sources.  
Figure~\ref{fig:radio colormag} demonstrates that GB6 detections are not
necessary to obtain an extensive characterization of these radio sources,
including radio spectral slope.  Sources as faint as 
$t_\mathrm{NVSS}\approx13.9$ can be studied without biasing toward steep-spectrum 
sources.  The sample presented here contains 63,660 radio sources (optical
subset: 18,728).  Constraining to $t_\mathrm{NVSS}<13.9$ results in a sample
of 50,046 (optical subset: 15,424).  This catalog represents a significant
increase in comprehensive radio data available to the scientific community. 

Figure~\ref{fig:radio colormag} provides a visualization of the unified radio
catalog in a five dimensional parameter space defined by four radio
magnitudes ($t_\mathrm{NVSS}, t_\mathrm{FIRST}$ (\emph{integrated}), and
$t_\mathrm{FIRST}$ (\emph{peak}) at 20\,cm; $t_\mathrm{WENSS}$ at 92\,cm)
and one optical morphology parameter (optical light concentration; see
\S\ref{subsubsec:sdss photometry}).  The four radio magnitudes are combined
into parameters indicating magnitude, morphology, and spectral slope.  Using
three radio morphology classes, radio magnitude, 
spectral slope, and three optical identification/morphology subclasses,
we have summarized the five-dimensional parameter space using the nine
two-dimensional panels (radio color-magnitude diagrams) in this figure.

The marginal distributions of spectral index $\alpha^{92}_{20}$ for the sample
in Figure~\ref{fig:radio colormag} are similar to the distributions shown in 
Figure~\ref{fig:alphamean all radio}.  However, the bimodality seen in the
compact sample is
reduced.  This is due both to the inclusion of fainter sources, and the
fact that one must use $\alpha^{92}_{20}$, rather than the average of
$\alpha^{92}_{20}$ and $\alpha^{20}_6$, to describe the radio spectra of a
sample that does not require detection at 6\,cm.

\begin{figure}
\figcaption{\label{fig:sdss matching statistics nfw}
Magnitude distributions for sources detected by NVSS, FIRST, and WENSS.
\emph{Panel~A:} Full sample divided by radio morphology.  
\emph{Panel~B:} Full sample divided by optical identification.
\emph{Panel~C:} Optically-identified sources divided by radio morphology.
\emph{Panels~D-F:} Complex, resolved, and compact radio sources,
respectively, divided by optical identification. }
\epsscale{0.85}
\plotone{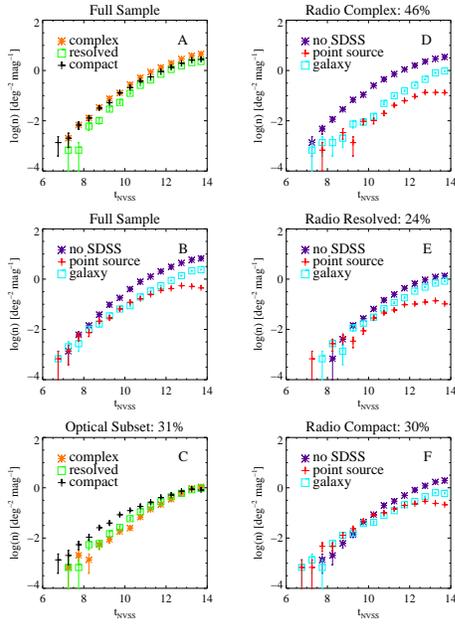}
\end{figure}

Magnitude distributions for the nine radio/optical morphology subclasses are
shown in Figure~\ref{fig:sdss matching statistics nfw}.  This is the equivalent
to Figure~\ref{fig:sdss matching statistics}, but corresponding to the five
times larger sample of NVSS-FIRST-WENSS sources.  This figure shows the
striking result that, at magnitudes fainter than $t_\mathrm{NVSS}=11.5$ 
($\approx$100\,mJy), the density of optical point sources flattens out
dramatically, rather than continuing to rise as would be expected for a
homogenous source distribution.  This catalog is the first sample of radio
sources both large and deep enough to investigate such faint source counts.
The same behavior is seen in all morphology classes: point source counts
flatten out while galaxy counts continue to rise at faint magnitudes. 

\subsection{Optical analysis aided by radio information}
\label{subsec:optical ids II}

In the previous section, we showed that analysis in radio space is greatly
aided by the addition of optical information.  Similarly, the standard optical
analysis, such as determination of optical luminosity function and its
evolution with redshift, can benefit from the addition of radio observations.
For example, one can ask whether radio loudness and radio spectral index
correlate with optical quantities (e.g., ``Can SDSS data alone predict the
strength and spectral properties of radio emission?'').  In this section, we
take initial steps in this direction by examining the dependence of radio
spectral slope on redshift and optical magnitude for quasars, and on optical
magnitude and color for galaxies.  We also investigate the relationship between
radio loudness, radio spectral slope, and optical color.

We adopt the spectral index $\alpha^{92}_i$ as a measure of radio
loudness, where
\begin{equation}
  \label{eq:alpha92i}
  \alpha^{92}_i=0.0657(t_\mathrm{WENSS}-i).
\end{equation}
This spectral index relates the 92\,cm flux to the SDSS $i$ band
flux\footnote{An alternative measure of radio loudness is based purely
  on radio luminosity.  \citet{i02} found that the two different quantities
  yield similar radio loudness classifications, owing to strong selection
  effects in flux-limited samples.}, and can be converted to the often used
radio loudness measure, $R_i = 0.4 (i-t_\mathrm{FIRST})$ \citep[e.g.][]{i02}
via
\begin{equation}
  \alpha^{92}_i = 0.11\alpha_{20}^{92} - 0.16\, R_i.
  \label{eq:convert}
\end{equation}                                           
A minimum in the radio loudness distribution at $R_i \sim 1$ reported
by \citet{i04qso} corresponds to
\begin{equation}
  \alpha^{92}_i = 0.11\alpha_{20}^{92} - 0.16 \sim -0.25.
\end{equation}
The NVSS-FIRST-WENSS sample analyzed here is not sufficiently
sensitive to investigate radio loudness bimodality, due to the
WENSS flux limit.                                       

Throughout this section, we focus on sources observed by the NVSS, FIRST,
WENSS, and SDSS surveys, with optical spectra from the SDSS.  We assume a
cosmology with $H_\mathrm{0} = 70~\mathrm{km~s}^{-1}\mathrm{Mpc}^{-1}$,
$\Omega_M=0.3$, and $\Omega_\Lambda=0.7$.

\subsubsection{Spectroscopically-identified quasars}
\label{subsubsec:spec quasars}

To find a sample of spectroscopically-identified radio quasars, we match
unified catalog sources to the Fifth Data Release (DR5) SDSS Quasar Catalog
\citep[][hereafter S07]{dr5quasars}.  This catalog is a cleaner quasar
sample than sources identified as quasars by the SDSS spectroscopic software
pipeline: the S07 spectra have been \emph{visually examined} to verify that
they have at least one broad emission line
(FWHM\,$\gtrsim1000~\mathrm{km~s}^{-1}$) or are unambiguously broadline 
quasars.  Candidates for the visual examination included all sources selected
by the SDSS quasar target selection algorithm, and all sources identified as 
quasars after spectroscopic observation. 

The sample used here consists of the 1,288 NVSS-FIRST-WENSS sources matched to
an S07 source within $2\arcsec$ (sample~H).  
Note that because the quasars were discovered via various SDSS target selection
algorithms, they do not represent a complete statistical dataset.  Indeed,
proximity to a FIRST source is one of the selection methods.  One therefore
must be 
careful when creating statistical samples from this data set; e.g. one must
choose quasars identified by the same target method, and account for sources in 
regions of the sky where different target selections were used.  Note however
that the SDSS optical quasar sample is estimated to be complete at the 94.6\%
level for the magnitude range $16<i<19.$ and redshifts less than 5.8
\citep{richards02}. 

\begin{figure}
\figcaption{\label{fig:quasar distributions}
Distributions in absolute magnitude-redshift-spectral index parameter space
for the spectroscopic quasar sample.
The dotted and dashed lines indicate the $i<20.2$ and $i<19.1$ limits,
respectively, used in the quasar target selection algorithm.  
\emph{Panel~A:} absolute magnitude vs. redshift for the full sample, with
bins in redshift and \emph{apparent} magnitude 
color-coded by the median value of $\alpha^{92}_{20}$ (\emph{upper colorbar}).
The histogram at the bottom of the panel shows the redshift distribution.
\emph{Panels~B through~D:} panel~A divided into
the three radio morphology subclasses.  The
percentage of quasars in each morphology class is given.  Full histograms show
the redshift distribution in each panel; for comparision the redshift
distribution of the full quasar sample is also shown (\emph{dotted line}).  Histograms
have been normalized such that all have equal area.
\emph{Panels~E and~F:} Spectral index vs. optical absolute magnitude
distributions for complex and compact radio sources, respectively.  Points are
color-coded by redshift.  Circles show the median value of $\alpha^{92}_{20}$
in bins of 1~magnitude.}
\epsscale{0.45}
\plotone{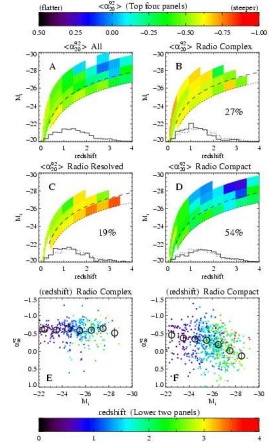}
\end{figure}

Figure~\ref{fig:quasar distributions} investigates the dependence of radio
spectral slope on optical luminosity and redshift for quasars with different
radio morphologies.  Optical K-corrections used to calculate absolute magnitude
$M_i$ assume an optical spectral index $\alpha=-0.5$ \citep{i02}.  Panel~A
shows that the radio spectral slope for radio quasars is independent of
redshift and is a strong function of absolute magnitude: \emph{quasars with the 
steepest radio spectra tend to be optically faint}.  We investigate this
correlation 
as a function of radio morphology in panels B-D: the trend is seen quite
strongly in the compact morphology class, but is weak for the complex and
resolved sources.  Additionally, compact quasars have the flattest radio
spectra, while radio spectra of complex sources are relatively steep.

\begin{figure}
\figcaption{\label{fig:quasar spec example}
A simulated radio spectrum for which $\alpha^{92}_{20}$ would appear flat at
high redshift (\emph{z}) and steep at low redshift.  Observed 20 and 92\,cm
fluxes correspond to shorter rest-frame wavelengths for a high-redshift 
object.}
\epsscale{0.5}
\plotone{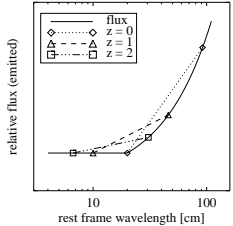}
\end{figure}

A more quantitative visualization of the correlation between radio spectral
slope and optical luminosity is shown in Panels~E and~F, for radio complex and 
radio compact sources, respectively.  The median value in unit magnitude bins
is indicated by black circles.  The median spectral slope for complex sources is
constant with luminosity, while the median slope for compact sources is a 
strong function of luminosity.  This trend may indicate a physical change in
radio spectral slope with optical luminosity, an orientation effect, and/or 
intrinsic variation in spectral slope combined with luminous sources being
observed at higher redshifts (i.e., a radio K-correction effect).  
Figure~\ref{fig:quasar spec example} shows as an example a simulated spectrum
that steepens 
for $\lambda > 20$\,cm, which would show the observed trend when viewed at
different redshifts.  It has been suggested however that spectral slope is an
indication of orientation \citep{jarvis},  and therefore of increased Doppler
beaming.  The spectrum of a quasar viewed along a line of sight close to the
jet axis is highly core-dominated.  It would thus appear flatter than a source
viewed off-axis, while beaming
would lead to a higher observed luminosity.  Complex sources (Panel~E) can be
expected not to show the same trend, as they almost certainly have extended
emission from steep-spectrum radio lobes, biasing against sources with
strongly-beamed jets.  On the other hand, the compact sample may contain
unresolved core-lobe sources, especially at higher redshift.  An angular
size of $5\arcsec$ (FIRST resolution) corresponds to $\sim40$\,kpc at
redshift~1.  According to \citet[][Fig.~2]{blundell}, a typical quasar can
reach this size in $\sim10^6-10^7$ years while the maximum active lifetime is
estimated to be $10^8-10^9$ years.  However, a line-of-sight alignment near the
jet axis would significantly shorten the projected linear size, decreasing the
likelihood of resolving a core-lobe source.  Note that the combination of
a flat core component and a steep lobe component can result in a spectrum of
the type shown in Figure~\ref{fig:quasar spec example}.  The trend observed in
Panel~F may therefore be due to an intertwined, and inseparable, combination of
the above effects.

Equivalent plots to Panels~E and~F (not shown) indicate that the median value
of $\alpha^{20}_6$ at all optical luminosities corresponds to a flat spectrum
between 6 and 20\,cm in all three morphology classes, which lends support to
the K-correction argument.  There are quasars whose spectra are consistent with
the simulated example shown in Figure~\ref{fig:quasar spec example}, but not
all, as 
demonstrated in Fig.~2 of \citet{kuehr}.  However, the resolution of their
observations was very low.  To accurately determine the radio spectral shape of
quasars will require better frequency resolution, which will be possible with
the future Expanded VLA project \citep{evla}.

\begin{figure}
\figcaption{\label{fig:quasar loudness}
Distribution of $\alpha^{92}_{ i}$, a measure of radio loudness, for the
spectroscopic quasar sample.  The top two panels show the region of radio
magnitude-optical magnitude space selected so as not to bias toward radio quiet
or radio loud sources.  The lower two panels show $\alpha^{92}_{ i}$ histograms
for red ($u-r>0.7$; \emph{solid line}) and blue ($u-r<0.7$; \emph{dashed line})
quasars.  The left panels correspond to sources with steep radio spectra;
panels on the right correspond to sources with flat radio spectra.}
\epsscale{1}
\plotone{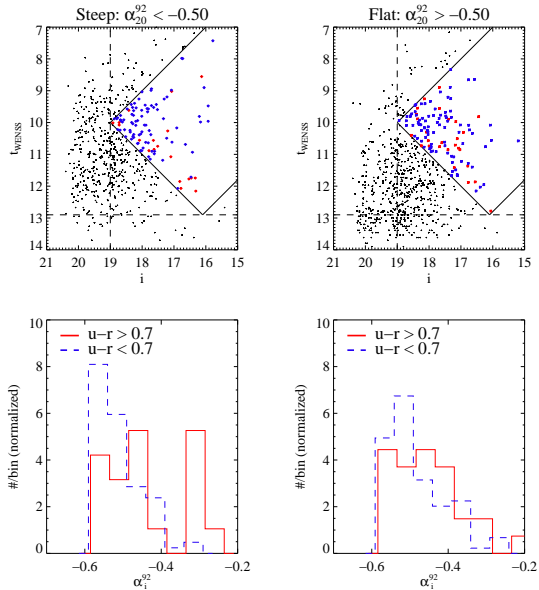}
\end{figure}

The dependence of radio loudness on radio spectral slope and optical color
for quasars is investigated in Figure~\ref{fig:quasar loudness}.  Sources are
divided into steep- (\emph{left}) and flat-spectrum (\emph{right}).  We require
$t_\mathrm{WENSS}<12.9$ so as not to bias against any sources with
$\alpha^{92}_{20}\gtrsim-1.5$.  We require $i<19$ to ensure a complete sample
of quasars.  Based on those two limits, we have selected a region of
radio/optical magnitude space which does not bias against radio quiet or radio
loud sources (\emph{top two panels}).  
The radio loudness distribution of steep-spectrum quasars (\emph{lower left
  panel}) is strongly dependent on color: a Kolmogorov-Smirnov test rejects the
null hypothesis that the two histograms share the same parent distribution at
the 99.6\% level.  No such dependence is seen for the flat-spectrum quasars 
(\emph{lower right panel}).  This paper presents the first observation of this
effect.  Note that the requirement of a detection
at 92\,cm biases these samples to bright radio fluxes, and prevents us from
reaching the expected local minimum in the $\alpha^{92}_i$ distribution at
$\alpha^{92}_i\sim -0.25$.

\subsubsection{Spectroscopically-identified galaxies}
\label{subsubsec:spec gals}

The sample of spectroscopically-identified galaxies (sample~G) is the set of
2,885 NVSS-FIRST-WENSS-SDSS matches whose spectra identify them as galaxies,
according to the SDSS automatic classification software.  Here, we restrict our
analysis to a flux-limited sample with $r<17.77$, in order to choose galaxies
from the SDSS main galaxy sample \citep{strauss}.  The resulting
flux-limited sample consists of 1,656 spectroscopic galaxies.

\begin{figure}
\figcaption{\label{fig:galaxy distributions morphology}
Distribution in color, magnitude, spectral index parameter space for the
flux-limited ($r<17.77$) spectroscopic galaxy sample.
\emph{Panel~A:} Color-magnitude diagram for the full sample with bins
color-coded by the median value of $\alpha^{92}_{20}$.  Contours show the
distribution of \emph{all} $r<17.77$ SDSS spectroscopic galaxies. 
The histogram below shows the $u-r$ distribution.
\emph{Panels~B through~D:} panel~A divided into the three radio morphology
classes, as labeled.  Points are color-coded by $\alpha^{92}_{20}$.  Percentage
of galaxies in the subclass is given in the lower right corner.  Dashed lines
at $u-r=3$ and $M_r=-22.5$ are provided to guide the eye.  The histogram
underneath shows the $u-r$ distribution (\emph{solid line}), compared to 
the distribution for the full sample (\emph{dotted line}; equivalent to solid line
histogram in Panel~A).}
\epsscale{0.85}
\plotone{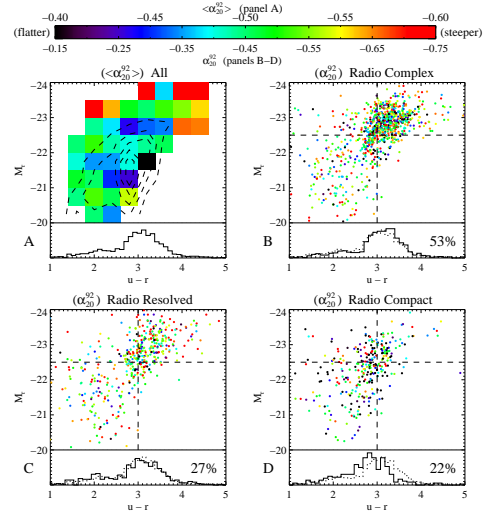}
\end{figure}

Figure~\ref{fig:galaxy distributions morphology} investigates the dependence of
radio spectral slope on optical luminosity and color for galaxies.  Complex
galaxies tend to have steeper radio slopes than compact galaxies.  Radio
sources with flat 
spectrum emission are generally understood to be quasar cores \citep[i.e., quasar
sources without radio lobes,][]{krolik}.  Quasar core sources are compact
in the radio due to their small size and typically large distances.  Sources in
panel D therefore beg the question: what are the flat-spectrum, radio compact
sources that are optically-identified as galaxies?  It is possible that those
sources are quasars too weak in the optical to overpower emission from
the galaxy, observed as galaxies with AGN (e.g. Seyfert galaxies, LINERS),
a possibility we will investigate in future work.

\begin{figure}
\figcaption{\label{fig:galaxy distributions slope}
Distributions in color, magnitude, and radio to optical spectral slope for
the flux-limited ($r<17.77$) spectroscopic galaxy sample.
\emph{Panels A and B:} Color-magnitude diagrams ($M_r$ vs. $u-r$)
for steep- and flat-spectrum radio sources, respectively; points are
color-coded by the radio to optical spectral slope ($\alpha^{92}_{ i}$).  
The solid line histogram shows the $u-r$ distribution; the dotted line
histogram shows the $u-r$ distribution of the full 
sample (equivalent to solid line histogram in Panel~A of Figure~\ref{fig:galaxy
  distributions morphology}).  The dashed vertical line indicates $u-r=2.5$.
The percentage of galaxies in each class is given in the lower right corner.
\emph{Panels C and D:} Normalized distributions of $\alpha^{92}_{ i}$ for red
($u-r>2.5$) and blue ($u-r<2.5$) radio galaxies from the top two panels.}
\epsscale{0.85}
\plotone{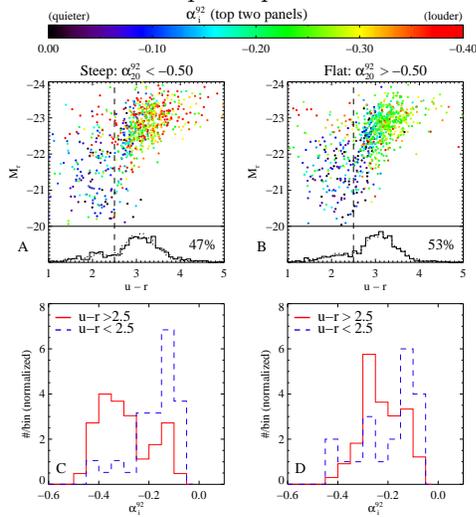}
\end{figure}

Figure~\ref{fig:galaxy distributions slope} explores the dependence of radio
loudness of galaxies on optical luminosity and color.  The top row shows the
distribution 
of steep- (\emph{left}) and flat-spectrum (\emph{right}) sources in optical
color-magnitude diagrams with points color-coded by $\alpha^{92}_i$.  Note
that only very luminous galaxies are radio loud ($\alpha^{92}_i\lesssim-0.35$
for $M_r<-22$).  The lower two panels show the distribution of
$\alpha^{92}_{~i}$ for steep and flat radio galaxies divided by $u-r$ color.
Similar to Fig.~\ref{fig:quasar loudness}, we focus on an unbiased region of 
radio/optical magnitude space, based on the limits $t_\mathrm{WENSS}<12.9$ and 
$i<17$.  The lower left panel shows a significant difference in radio loudness
for blue (\emph{dashed line}) and red (\emph{solid line}) steep-spectrum radio
galaxies.  The blue galaxies have a median $\alpha^{92}_i$ of $\sim-0.13$,
while the median value for red galaxies is $\sim-0.26$.  A difference of 0.13
in $\alpha^{92}_i$ (Eq.~\ref{eq:alpha92i}) corresponds to a change of about 2
magnitudes, i.e. steep-spectrum red galaxies are about two orders of magnitude
brighter in the radio than steep-spectrum blue galaxies of the same optical
brightness.  The distributions of radio loudness for blue and red flat-spectrum
galaxies are not significantly different.

\section{CATALOG APPLICATIONS/FUTURE WORK}
\label{sec:discussion}

The information provided by this catalog will enable many diverse studies.  In
this section, we briefly remark on applications toward the investigation of the 
radio quasar/galaxy unification theory, the search for radio stars, and the
selection of possible high-redshift galaxies.  We then describe the usefulness
of the catalog for studies of the evolution of the radio universe by comparison
with models, which we discuss in detail in a companion paper (A. Kimball et
al. in preparation).

\subsection{Unification paradigm for radio-loud active galactic nuclei}

The orientation-based unification scenario \citep{urry_review} was originally
motivated by the similarity of radio emission from sources that appear very  
different optically, i.e. quasars and galaxies.  The theory assumes that
sources whose emission is dominated by a radio core or lobes are members of the 
same ``parent'' population, but differ in appearance because their highly 
anisotropic emission is viewed from different angles.  This anisotropy is
predominantly due to relativistic motion of the plasma in the inner jets and
Doppler boosting (``beaming'') when the angle between the line of sight and the
plasma velocity vector is small \citep{krolik}.  Hence, the same object could
appear as a core source when viewed at small angles (with the ``boosted'' core
outshining the lobes), and as an extended double-lobe or a core-lobe source
otherwise.  Understanding the radiation anisotropies in AGNs is required to
unify the different types; that is, to identify each single, underlying AGN
type that gives rise to different observed classes through different
orientations \citep{antonucci,urry_review}. 

The unified catalog allows users to classify many quasars and galaxies
according to their radio morphology (e.g., core sources, lobe sources) and
spectral behavior.  Statistical studies of such objects will lend evidence for
or against the unification paradigm by the investigation of number statistics,
the size versus orientation distribution, and environment.  The unification 
scenario has successfully explained many observed properties of bright radio
sources in previous research \citep{barthel,urry,padovani,lister,urry_review},
but recent studies suggest there may be some intrinsic differences between
radio quasars and galaxies \citep{willott02}.

\subsection{A method for selecting high-redshift ($z>1$) galaxy candidates}
\label{subsec:highz gals}

Galaxies at high redshift are important for studies of large scale structure
and galaxy evolution, and in fact the higher the redshift of a galaxy the more
useful it is for these studies.  Candidate high-redshift galaxies can be
selected from the unified catalog based on their radio morphology, radio
spectral slope, and lack of an optical counterpart.  The lack of optical
counterpart requirement is intended to select sources so distant that their
observed optical emission lies blueward of the rest-frame $4000$\AA ~break
\citep[e.g., ][]{madau}.  As the optical sources used herein were selected in
the $r$ and $i$ bands at 6165\AA ~and 
7481\AA ~respectively, this requirement will tend to select galaxies at redshifts
of $z\sim1$ and higher.  A steep radio spectrum has often
been used as a criterion to find high-redshift sources, owing to the fact that
higher redshift objects are observed at higher rest-frame frequencies, and
because radio galaxy spectra tend to flatten below $\sim300$\,MHz in their rest
frame \citep[e.g.,][and references therein]{cruz}.
Additionally, as discussed
in \S\ref{subsec:optical ids I}, in the compact morphology subclass a steep
radio spectrum is a likely indicator of an optically-resolved source.

To find a sample of candidate high-redshift galaxies, we select sources
which are unresolved by FIRST, undetected by the SDSS, and have steep radio
spectra.  Specifically,
we begin with the sample of objects identified by NVSS, FIRST, and WENSS which 
have compact radio morphology.  We require $\alpha^{92}_{20}<-0.5$ and no SDSS
match within $3\arcsec$.  Note that
heavily dust-obscured galaxies are also likely to be found in such a sample.
In the catalog overlap region, the above criteria select 9,953 sources.
Visual examination of SDSS images of all the candidates revealed 334 suspect
sources, i.e. radio positions coinciding with a diffraction spike, scattered
light, a nearby star or galaxy, etc.  Removing the suspicious sources results
in a sample of 9,619 candidate $z>1$ galaxies.  It will be
interesting to compare this sample with deeper optical surveys and observations
at other wavelengths, as the data become available.  This sample is
available for download on the catalog website (sample~K).

\subsection{The search for radio stars}
\label{sec:radiostars}

With the great increase in the sensitivity of radio surveys in the last several
decades, along with the more accurate source positions afforded by radio
interferometry, comes the ability to search for fainter and rarer radio sources.
A small fraction of stars have significant non-thermal radio emission, such as
dMe flare stars \citep{white} and cataclysmic variables
\citep{chanmugam,mason}, but most stars have only weak thermal radio emission.
Because quasars and stars lie in different locations in SDSS color-color
diagrams \citep{i02,richards01}, it may be possible to select a clean sample of
radio stars based on their photometric colors.  To investigate this
possibility, we select a sample of radio star candidates from the unified
radio catalog.

We begin by selecting point sources from the catalog overlap region which
coincide with a FIRST radio detection.  We use a conservative matching distance
of $1\arcsec$ to limit the contamination due to random matches.  To ensure a
sample for which the majority have spectra, we require $i<19$.\footnote{The SDSS 
  quasar target selection algorithm flags all $i<19.1$ sources within
  $2\arcsec$ of a FIRST detection \citep{dr5quasars}.}  We eliminate the region
of color-color space where quasars are commonly found: $u-g<0.8$ and
$-0.2<g-r<0.6$.

The number of candidates remaining after each step of the selection algorithm
is shown in Table~\ref{table:radiostars}.  The resulting sample of 532
candidate radio stars contains 406 objects with SDSS spectra.\footnote{The
  remaining 126 stellar candidates do not have spectra for one or more of the
  following reasons: they are not in the region of sky included in the SDSS DR6
  spectroscopic sample, they are too close on the sky to another spectral
  target, they failed the quasar target selection algorithm
  because of the bright limit or bad photometry, or they were not targeted
  because the photometry in a previous SDSS data release led to an extended
  rather than point source optical morphology identification.}
From the SDSS spectral classifications,\footnote{Spectroscopic identifications
  determined using the SDSS ``SpecBS'' Princeton reductions.  For details, see
  http://spectro.princeton.edu/.} we determine that 78 are
stars, 313 are quasars, and 15 are galaxies.  The results of applying the
selection to the SDSS DR5 quasar catalog \citep{dr5quasars} are also shown for
comparison.  The completeness of the quasar catalog is $\sim95\%$
\citep{richards02,vandenberk}.  Seven of the known quasars were identified as
galaxies by the SDSS spectroscopic pipeline, and two as stars.  The latter are
actually the super-position of a star and a quasar: the spectra are primarily
stellar, but also contain at least one broad emission line.  Note that 33 of
the candidate radio sources classified as ``quasar'' are absent from
the quasar catalog.  These are sources that were spectrocopically-observed for
the sixth SDSS data release, but are not in DR5.

The sky density of unresolved $i<19$ SDSS sources is $\sim1000\deg^{-2}$, while
the density of quasars is $\sim10\deg^{-2}$.  However, out of nearly 6 million
point 
sources with $i<19$, we found only 78 confirmed radio stars, implying that the
likelihood of a star being a radio source is approximately $10^{-5}$, while the
likelihood for a quasar is about $8\times10^{-2}$.  Consequently, despite using
a conservative color cut to eliminate probable quasars, the sample is
dominated by radio quasars rather than radio stars.  Using this color-color
selection is therefore not an efficient way to find radio stars.  One must have
spectral classification in order to select a clean radio star sample;
photometric observations alone are not sufficient.

\subsection{Radio galaxy evolution models}

In the companion paper, we attempt to understand and quantitatively explain
the observed source counts for the nine subclasses defined by radio morphology
and optical identification (Fig.~\ref{fig:radio colorcolor optical}), with the
aid of models.  \citet{barai06,barai07} investigated a large number of input
parameter variations for several radio galaxy evolution models and compared the
resulting distributions of radio power, spectral index, linear size, and
redshift to the low-frequency 
($\sim$2\,m) Cambridge catalogs \citep{laing,eales,mcgilchrist}.  They found
that none of the models was able to provide good simultaneous fits to all
observables for all three catalogs, but noted that the observational
datasets were too small to adequately constrain the model parameters. 

The catalog presented in this paper represents a significant increase in the
size of radio samples for which luminosity, size, spectral slope, and redshift
can be measured.  We will compare the observed distributions from the unified
catalog to distributions of sources from mock catalogs (obtained by applying
observational selection criteria to the computer-generated radio skies).
The distributions will help to constrain the increasingly sophisticated models
of radio galaxy evolution.  The models will thus help explain the physics and 
evolution leading to the distributions seen in the unified catalog. 

\subsubsection{Defining a long-wavelength flux-limited sample}
\label{subsubsec:wenss sample}

The radio galaxy evolution models currently perform best at long wavelengths,
where emission is dominated by the radio lobes.  The most robust approach for
comparing models and observations is therefore to define a flux-limited sample
using long-wavelength measurements, such as the 92\,cm flux from the WENSS
survey.  Fortunately, because most radio sources are brighter at longer radio
wavelengths, a source observed in WENSS (faint limit of 18\,mJy) is almost
guaranteed to be detected at 20\,cm (faint limits of 1\,mJy for FIRST and
2.5\,mJy for NVSS).  However, the 92\,cm flux limit must be chosen carefully in
order to avoid biasing against sources with radio spectra so steep that they
are too faint at 20\,cm to be detected by NVSS.  The color-magnitude diagrams
and spectral slope histograms presented in the data analysis of this paper
indicate that very few radio sources have a spectral slope index less than -1.5 
between 20 and 92\,cm.  To ensure an NVSS detection for all sources with
$\alpha^{92}_{20}<-1.5$ in a flux-limited sample requires
$t_\mathrm{WENSS}\leq12.9$ (see Eq.~\ref{eq:alpha9220}).  We positionally
matched all $t_\mathrm{WENSS}<12.9$ sources in the WENSS survey to NVSS and
FIRST, and found that only 1\% did not have a match in NVSS.  We interpret this
value as an upper limit on the fraction of sources with
$\alpha^{92}_{20}<-1.5$.  Ninety-five per cent of the WENSS sources have both a
FIRST detection and an NVSS detection within $30\arcsec$.  The sample of
NVSS-FIRST-WENSS matches with $t_\mathrm{WENSS}<12.9$ in the unified catalog
contains 47,567 sources.  Over 12,000 of those have optical counterparts, from
which a cosmological distance can be determined via photometric redshift
techniques.  Photometric redshifts from the SDSS are accurate to $\sim0.06$ for
luminous red galaxies with redshifts $z<0.7$ \citep{nikhil}.  A further 3,000
of the radio-optical sample additionally have SDSS spectroscopic redshifts.  This 
dataset therefore represents a major size increase compared to the sample of
327 sources used by Barai \& Wiita.

\section{SUMMARY}

We have presented a unified catalog of radio objects by combining observations
from several radio and optical surveys.  The complete catalog is available
online, along with scientifically useful data subsets.  Using several
parameters determined from the measured radio fluxes, we investigated the
distribution of sources in radio color-magnitude-morphology space and optical
color-redshift-luminosity space.

To create the catalog, we merged together the two 20\,cm radio surveys, FIRST
and NVSS.  Included are FIRST-NVSS associations (three sources from each survey
with the closest proximity to sources in the other survey, within $30\arcsec$),
FIRST sources with no NVSS counterpart, and NVSS sources without a
nearby FIRST counterpart.  Where sources had a counterpart in one of the other
three surveys, we supplemented the catalog data with 6\,cm, 92\,cm, and optical
observations. 

For a preliminary data analysis, we converted the radio fluxes (6\,cm, peak
20\,cm, high- and low-resolution integrated 20\,cm, and 92\,cm) into several
useful parameters describing 20\,cm radio magnitude, 20\,cm radio morphology,
radio spectral slope, and optical identification/luminosity.  The automatic
morphology classification divides sources into ``complex'', ``resolved'',
and ``compact'' categories, with results that compare well to visual morphology
classification.  Using various radio color-magnitude diagrams, we found that
most radio sources have a steep radio spectral slope between 6 and 20\,cm
($\alpha\sim-0.8$, where $F_\nu\propto\nu^\alpha$).  However, the radio
compact sources contain a significant fraction of flat-spectrum sources
($\alpha\sim0$).   SDSS morphological identifications for optically-matched
sources indicate that the steep-spectrum compact sources are primarily
galaxies, while flat-spectrum compact sources are primarily optically
unresolved (i.e., quasars).  The complex and radio resolved sources
contain comparable numbers of both galaxies and optical point sources.

Focusing on sources with SDSS spectra, we investigated the dependence of radio
spectral slope and radio loudness on optical quantities (color, magnitude,
redshift) for quasars and galaxies.  For complex and radio resolved quasars,
no significant dependence of spectral slope on redshift or optical magnitude
was found.  For compact quasars, however, radio spectral slope seems to flatten
out with increasing redshift/luminosity.  The trend is consistent with a radio
K-correction resulting from a spectrum that is steep at wavelengths longer than
20\,cm and flat for wavelengths shorter than 20\,cm, but may also be an effect
of orientation and Doppler beaming.  Formally, a
Kolmogorov-Smirnov test showed a dependence on optical color for the radio
loudness distribution of steep-spectrum quasars.  No color-dependence was found
for the flat-spectrum quasars.
In the galaxy sample, the compact morphology class tends to be bluer and
fainter, with flatter radio slopes, than either the complex or radio resolved
class.  Among steep-radio-spectrum galaxies, the red sources have radio fluxes
on average two orders of magnitude brighter than the blue sample. 

We have presented methods for selecting candidate radio stars and candidate
high-redshift galaxies from the unified catalog.  We concluded that a
conservative color-selection attempting to reduce contamination from quasars
still yields a sample that is completely dominated by quasars rather than radio
stars.  Spectra are therefore necessary to successfully identify radio stars.  

In a companion paper (A. Kimball et al. in preparation), we will compare the
results of several 
state-of-the-art models for radio galaxy evolution to observations yielded by
the unified catalog, and test the unification paradigm for powerful radio
galaxies and quasars.

While this is not the first catalog to combine observations in the radio and
the optical to further the study of radio galaxies and radio quasars, the
unified catalog presented in this paper has significant advantages over
previous studies.  It is the largest catalog to date that combines optical
observations with multi-wavelength radio data.  It can be used to define a
long-wavelength (92\,cm) flux-limited sample of nearly 50,000 sources detected
in WENSS, FIRST, and NVSS.  Comparing 20\,cm FIRST and NVSS fluxes allows us to
place every one of those sources into one of the three radio morphology
classes, and the additional 92\,cm detection allows us to calculate a radio
spectral slope.  Over one quarter of the long-wavelength flux-limited
sample has an optical detection in the SDSS, including over 3,000 with spectra.
The usefulness of the catalog will increase as larger surveys at different
wavelengths with deeper observations become available.  For example, the deep
observations planned for the future Large Synoptic Survey
Telescope\footnote{\emph{www.lsst.org}} will provide a nearly 100\% optical
identification rate, and extremely accurate photometry in six bands.

\acknowledgements

This material is based upon work supported under a National Science Foundation
Graduate Research Fellowship, and by NSF grant AST-0507259 to the University of
Washington.  The authors would also like to thank Wim de Vries and Paul Wiita
for their helpful comments.

\appendix

\section{CATALOG MATCHING}
\label{app:matching}

\subsection{Matching FIRST and NVSS}

In this section, we describe in detail the process of matching sources in
FIRST and NVSS, and present a method for selecting a clean sample of 20\,cm
sources.

Before proceeding, it is necessary to define certain clarifying terms, some of
which are data parameters found in the catalog.  Our matching algorithm used a
\emph{primary} survey and a \emph{secondary} survey.
Source positions in the primary survey were used
as search centers, and neighbors within a certain radius were found in the
secondary survey.  To find catalog sources we applied the matching algorithm
twice: with FIRST as primary and NVSS as secondary, and again with NVSS as
primary and FIRST as secondary.  Each time,
the three closest neighbors from the secondary survey were ranked by proximity
to the primary source (within the pre-determined search radius).
The advantage of performing the matching twice is that we can recover
the three closest FIRST detections to a NVSS primary, and vice-versa. 
Each catalog row contains records for a single NVSS source and a single FIRST
source.  For an entry with an NVSS primary, the parameter
\emph{matchflag\_nvss} is set to $-1$, while \emph{matchflag\_first} is set
either to 0, indicating a primary source with no FIRST match, or to
1, 2, or 3 indicating the proximity ranking of the FIRST match.
Conversely, for FIRST primaries \emph{matchflag\_first} is set
to\,$-1$, and \emph{matchflag\_nvss} equals 0, 1, 2, or~3.
Note that a single source can appear
as the primary object in as many as three separate catalog entries.

\begin{figure}
\figcaption{\label{fig:cartoon}
Examples of different types of sources found by matching NVSS and FIRST.
Circles represent NVSS detections; squares represent FIRST detections.  The
sizes of the two types of symbols help illustrate the different spatial
resolution of the two surveys.  Quadrant~A contains a complex source
resolved into three components by FIRST but unresolved by NVSS, as well as a
fourth (unrelated) FIRST detection.  Quadrant~B contains a simple source
identified by both surveys, with no nearby neighbors.  Quadrant~C contains a
FIRST detection with no NVSS counterpart, while Quadrant~D contains a NVSS
detection with no FIRST counterpart.}
\epsscale{1}
\plotone{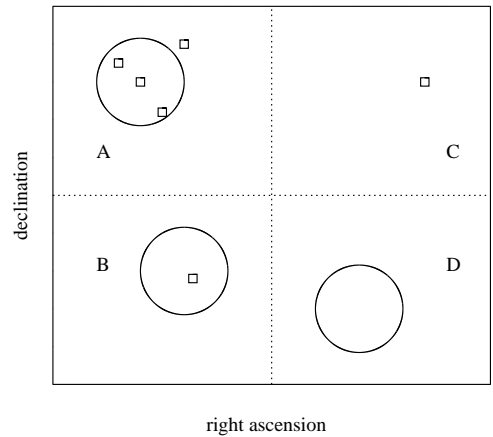}
\end{figure}

Due to the lower spatial resolution and higher flux limit of NVSS, 
individual NVSS sources can be resolved into separate components by FIRST.  
Therefore, multiple FIRST
detections matched to the same NVSS primary are common, while multiple NVSS
detections matched to the same FIRST primary are rare.  
Figure~\ref{fig:cartoon} gives a visual example of common types of objects
found in the catalog.  FIRST sources are shown as small squares and NVSS
sources as large circles.  In quadrant~A, an unresolved NVSS source is resolved
into three components in FIRST (example: a radio quasar with two lobes); a
fourth FIRST detection lies nearby.  When taking FIRST as the primary survey,
each of the four FIRST sources would identify the NVSS object as the nearest
neighbor.  When taking NVSS as the primary survey, the three closest FIRST
sources would match to the NVSS source while the fourth would not.
The NVSS source in Quadrant~A would appear in \emph{seven} catalog entries.
It would appear in four rows with \emph{matchflag\_first}\,$=-1$ and
\emph{matchflag\_nvss}\,=\,1.  It would appear in three rows with
\emph{matchflag\_nvss}\,$=-1$ and \emph{matchflag\_first}\,=\,1, 2, or~3.  To cull
out the duplicates in order to select a clean sample, one would simply require
\emph{matchflag\_nvss}\,$=-1$ and \emph{matchflag\_first}\,$=1$.

The remaining objects in Figure~\ref{fig:cartoon} are simpler to match.
Quadrant~B contains a simple, isolated source identified by both surveys
(examples: a radio quasar with no jets or a distant galaxy).
Such an object would appear in \emph{two} catalog rows: once with
\emph{matchflag\_first}\,$=-1$ and \emph{matchflag\_nvss}\,=\,1, and a second
time with \emph{matchflag\_nvss}\,$=-1$ and \emph{matchflag\_first}\,=\,1.
Quadrant~C contains a 
FIRST source without a NVSS counterpart
(e.g., an object too faint to be detected by the latter);
it would appear once in the catalog, with \emph{matchflag\_first}\,$=-1$ and
\emph{matchflag\_nvss}\,=\,0. 
In Quadrant~D is a NVSS source unmatched by FIRST (e.g., 
a low surface-brightness galaxy).  This source's single
catalog entry would have \emph{matchflag\_nvss}\,$=-1$ and
\emph{matchflag\_first}\,=\,0. 

To find all sources detected by both surveys while minimizing
duplicates, the simplest selection is to require \emph{matchflag\_nvss}\,$=-1$ and
\emph{matchflag\_first}\,=\,1.  The resulting data set consists of 142,622 sources
in the overlap region, with only 158 (0.11\%) duplicate NVSS sources matched to
the same FIRST source.  Specific examples of data selections are given in
Appendix~\ref{app:data samples}. 

\subsection{Matching to the SDSS}

To match to the SDSS photometric catalog, we implemented a different
technique from that used in matching the radio surveys.  The motivation was to 
find the nearest neighbor, as well as the brightest neighbor within a
pre-defined 
distance $D$, where $D$ was adjusted according to the distance of the nearest
neighbor.  We began by finding the nearest optical neighbor
within $60\arcsec$ of each catalog entry.  Where the distance $d_{near}$ to the
nearest neighbor was less than $3\arcsec$, we defined $D=3\arcsec$.  For
$3\arcsec<d_{near}<10\arcsec$, we defined $D=10\arcsec$, for
$10\arcsec<d_{near}<30\arcsec$ we defined $D=30\arcsec$, and for
$30\arcsec<d_{near}<60\arcsec$ we defined $D=60\arcsec$.  
We then recorded the \emph{brightest} neighbor within distance~$D$.  
Each catalog entry contains two sets of data parameters corresponding to SDSS
photometric matches: one set is identified by the prefix
``near'' and the other by the prefix ``bright''
(e.g. ``\emph{near\_modelmag\_u}'', 
``\emph{bright\_modelmag\_u}'').  Thirty-five per cent of FIRST sources in the
overlap region have an SDSS counterpart within $3\arcsec$, 61\% within
$10\arcsec$, and over 98\% within $30\arcsec$.  In 98.8\% of cases where an
optical match was found within $3\arcsec$, the brightest and the nearest SDSS
objects within $3\arcsec$ are the same.

\section{SELECTING DATA SAMPLES}
\label{app:data samples}

In this section, we define in detail the samples described in
this paper, and present several scientifically-useful data subsets
available for download on the catalog
website.\footnote{\emph{http://www.astro.washington.edu/akimball/radiocat/}}
Samples described in this appendix but not available online can be assembled
from the complete catalog using the selection requirements given here. 

Table~\ref{table:data} presents a summary of the data sets and their selection
requirements.  The meanings of catalog parameters referenced in the table
(in \emph{italics}) are as follows.  Parameters \emph{ra} and \emph{dec} are
right ascension and declination, respectively, in decimal degrees. 
\emph{Overlap} indicates a source inside (value~1) or outside (value~0) the
catalog overlap region.
\emph{Matchflag\_nvss} and \emph{matchflag\_first} are set to $-1$ to
indicate primary source catalog, 0 to indicate an isolated source, or 1, 2, 3 to
rank matched source by proximity to primary source (see
Appendix~\ref{app:matching} for details).  \emph{Distance} indicates separation
between FIRST and NVSS neighbors, in arcseconds.  \emph{First\_id},
\emph{wenss\_id} and \emph{gb6\_id} are unique identifiers for the FIRST, WENSS
and GB6 sources, respectively. 
\emph{Wenss\_distance} and \emph{gb6\_distance} indicate separation, in
arcseconds, between WENSS or GB6 neighbors and the 20\,cm source.
\emph{Near\_distance} and \emph{near\_type} indicate respectively the distance
to and object type of the nearest neighbor in the SDSS photometric survey, while
\emph{spec\_type} indicates the SDSS spectral classification for objects from
the spectroscopic survey.
\emph{Matchtot30} indicates the number of neighbors within $30\arcsec$ of the
primary 20\,cm survey source (where neighbors come from the secondary 20\,cm
survey).  For a full description of all 117 catalog parameters, we refer the
reader to the catalog website.

\subsection{Data sets discussed in the paper}

The data samples discussed in this paper consist of sources detected by both
FIRST and NVSS, sources detected by the three largest radio surveys (FIRST,
NVSS, WENSS), sources detected by all four radio surveys (FIRST, NVSS, WENSS,
GB6), and sources detected by all four radio surveys as well as the SDSS.
Spectroscopic galaxy and quasar samples are created from the FIRST-NVSS-WENSS
sources. 

Sources detected by both FIRST and NVSS, matched within $25\arcsec$, are listed
as sample~C in Table~\ref{table:data}.  
An efficient way of finding these pairs while culling duplicates is to select
NVSS sources, and their nearest FIRST neighbor.
These
sources can be selected by requiring 
\emph{overlap}\,=\,1, \emph{matchflag\_nvss}\,=$-1$,
\emph{matchflag\_first}\,=\,1, and 
\emph{distance}\,$\leq25$.  The resulting sample contains 141,881 objects
($48\deg^{-2}$) detected by both 20\,cm surveys, which can therefore be
classified by radio morphology.

Sample~D contains sources detected by FIRST, NVSS, and WENSS.  This is the
subset of sample~C with a WENSS counterpart within $30\arcsec$.  Additional
requirements to select this data set are \emph{wenss\_id}\,$\neq0$ and
\emph{wenss\_distance}\,$\leq30$.  The resulting sample contains 63,660 sources
($21.5\deg^{-2}$).
These sources have three 20\,cm fluxes and a 92\,cm flux measurement, and can
therefore be categorized by both radio morphology and spectral slope
($\alpha^{92}_{20}$). 
 
Sample~E contains sources detected by all four radio surveys; it is
the subset of sample~D that has a GB6 counterpart within $70\arcsec$.  Selecting
these sources requires additional parameter settings \emph{gb6\_id}\,$\neq0$ and
\emph{gb6\_distance}\,$\leq70$.
The sample is comprised of 12,414 sources ($4.2\deg^{-2}$), which can be
categorized by radio morphology and two spectral slope measures.

Sample~F contains sources detected by all four radio surveys and the SDSS.
To select these sources from sample~E requires \emph{near\_type}\,$\neq0$ and
\emph{near\_distance}\,$\leq2$.  The sample contains 4,732 sources
($1.6\deg^{-2}$), 
which can be categorized by radio morphology and two spectral slope measures,
and identified as galaxies or point sources.

Sample~G contains spectroscopically-identified galaxies detected by NVSS,
FIRST, WENSS, and SDSS.  These sources are a subset of Sample~D, requiring 
\emph{spec\_type}\,=\,2 and \emph{near\_distance}\,$\leq2$.  It contains 2,885
galaxies.  Note that in \S\ref{subsubsec:spec gals}, we
restricted our analysis to a smaller, flux-limited sample of 1,656
spectroscopic galaxies with $r<17.77$.

Defining the spectroscopic quasar sample was more complicated than defining the
spectroscopic galaxy sample.  As discussed in \S\ref{subsubsec:sdss spectra},
the quasar catalog of \citet{dr5quasars} is cleaner and hence more reliable
than the set of spectroscopically-identified quasars supplied by SDSS DR6.  
To identify radio quasars with spectra, we therefore opted to match radio
sources to the \citet{dr5quasars} catalog, temporarily ignoring the
spectroscopic identifications obtained from SDSS DR6.  The sample we analyzed
(sample~H) consists of 1,288 NVSS-FIRST-WENSS sources with a spectroscopic
quasar within $2\arcsec$.  This file is available only in ASCII format with
105 columns, and contains data not available in the full catalog.  The first 74
columns contain data specific to the DR5 known quasar catalog; the remaining 31
rows contain radio fluxes and other data parameters from the unified catalog.
A header in the file gives a description of all 105 columns.

\subsection{Other data sets available online}

Other data files available online provide the complete catalog, a small spatial
subset containing all data parameters, isolated FIRST-NVSS sources, 
isolated FIRST-NVSS sources with a counterpart in WENSS or GB6, and isolated
FIRST-NVSS sources with optical counterparts.

The complete catalog (sample~A) is available as a compressed \emph{tar}
archive.  It contains 72 files (in \emph{fits} format), each covering a strip
of sky with a width of 5\degr in right ascension.  This archive contains all
117 data parameters for each of the 2,724,343 catalog entries; no selections of
any kind were applied to this sample.  A full explanation of all the data
parameters can be found on the website.

A small subset (sample~B) of the catalog is also available, intended for
testing purposes, to demonstrate the catalog format.  It covers approximately
$106\deg^2$ of sky in the range $150\degr<$~R.A.\,$<165\degr$ and
$40\degr<$~dec.\,$<50\degr$.  The file contains all 117 data parameters for the
16,453 catalog entries in that sky region.

We provide a set of isolated FIRST-NVSS matches within the
catalog overlap region (sample~I).
The set of isolated sources is comprised of 109,825 NVSS-FIRST detections
within a matching radius of $15\arcsec$, and no other FIRST neighbors within
$30\arcsec$.  The file includes positions (right ascension and declination of
the FIRST source), distance between the FIRST and NVSS positions, and 20\,cm
fluxes.  For sources with a match in GB6 or WENSS within $120\arcsec$ (47,490
total), it also contains distance to the matched source along with its flux.
Sources without a match have the corresponding data
parameters equal to zero.

We provide the subset of isolated FIRST-NVSS matches with a SDSS counterpart
within $2\arcsec$ (sample~J).  There are 44,851 such sources in the catalog
overlap region.  In addition to data parameters describing FIRST and NVSS
distance and fluxes, this file contains the position, photometric type,
magnitudes, and magnitude errors for the nearest SDSS match.

Finally, we provide the set of candidate 9,619 high-redshift galaxies discussed
in \S\ref{subsec:highz gals} (sample~K). 

\clearpage
\bibliography{unified_catalog}

\clearpage

\clearpage

\clearpage

\clearpage

\begin{deluxetable}{lcccc} 
\tablewidth{0in}
\tablecaption{\label{table:surveys}
Technical comparison of the radio source surveys.}
\tablehead{\colhead{statistics} & \colhead{FIRST} & \colhead{NVSS} &
  \colhead{GB6} & \colhead{WENSS}} 
\startdata 
Frequency [MHz] & 1400 & 1400 & 5000 & 326 \\ 
Wavelength [cm] & 20 & 20 & 6 & 92 \\ 
Resolution [arcsec] & 5 & 45 & 210 & 54 \\ 
Sky coverage & $b>30\degr$ & $\delta>-40\degr$ & $0<\delta<75$ & $\delta>30\degr$ \\
Sky area [$\deg^{-2}$] & 10,000 & 33,000 & 17,000 & 10,000 \\ 
Lim. Flux density & 1 & 2.5 & 18 & 18 \\ 
($5\sigma_{rms}$, mJy) \\ 
Lim. AB mag & 16.4 & 15.4 & 13.3 & 13.3 \\ 
Source density [$\deg^{-2}$] & 97 & 55 & 4.4 & 22 \\
\enddata 
\end{deluxetable}

\begin{deluxetable}{lrrr}
\tablewidth{0in}
\tablecaption{\label{table:matching radii}
Completeness and efficiency as a function of matching radius}
\tablehead{
\colhead{Matched surveys} & \colhead{Radius [arcsec]} &
\colhead{Completeness} & \colhead{Efficiency}} 
\startdata
FIRST-NVSS & 25 & 0.990 & 0.96 \\
& 30 & 0.997 & 0.96 \\
FIRST-WENSS & 30 & 0.990 & 0.92 \\
& 120 & 1.000 & 0.74 \\
FIRST-GB6 & 70 & 0.979 & 0.79 \\
& 120 & 1.000 & 0.68 \\
FIRST-SDSS & 2 & 0.982 & 0.95 \\
& 3 & 0.998 & 0.90 \\
& 10 & 1.000 & 0.47 \\
NVSS-WENSS & 55 & 0.910 & 0.85 \\
& 120 & 0.995 & 0.57 \\
\enddata
\tablecomments{To calculate these values, the nearest neighbor 
distributions were fit to a Gaussian (representing physical matches) plus a
rising linear function (representing random matches).  
}
\end{deluxetable}

\begin{deluxetable}{rrrrrrr}
\tablewidth{0in}
\tablecaption{\label{table:gaussfits}
Double-Gaussian fits to FIRST-NVSS $\Delta t$ distributions} 
\tablehead{
\colhead{Bin} & \multicolumn{3}{c}{Simple sources} & 
\multicolumn{3}{c}{Complex sources}\\
\colhead{} & \colhead{Fraction\tablenotemark{a}} &
\colhead{Mean\tablenotemark{b}} & \colhead{$\sigma$\tablenotemark{c}} &
\colhead{Fraction\tablenotemark{a}} & \colhead{Mean\tablenotemark{b}} &
\colhead{$\sigma$\tablenotemark{c}} }
\startdata
       $t_\mathrm{NVSS}<12$ & 0.59 & 0.035 & 0.044 & 0.41 & 0.58 & 0.35 \\
$12 < t_\mathrm{NVSS} < 13$ & 0.67 & 0.064 & 0.074 & 0.33 & 0.64 & 0.37 \\
$13 < t_\mathrm{NVSS} < 14$ & 0.65 & 0.12  & 0.13  & 0.35 & 0.75 & 0.47 \\
$14 < t_\mathrm{NVSS} < 15$ & 0.59 & 0.23  & 0.24  & 0.41 & 0.75 & 0.49 \\
\enddata
\tablecomments{The distributions are shown in the lower panel of
  Figure~\ref{fig:I02 fig13}.}
\tablenotetext{a}{Fraction of total area belonging to each Gaussian.}
\tablenotetext{b}{Position of Gaussian mean in $\Delta t$.}
\tablenotetext{c}{Standard deviation.}
\end{deluxetable}

\begin{deluxetable}{lrrrrrrrr}
\tablewidth{0in}
\tablecaption{\label{table:morphology}
Comparison of automatic and visual morphology classification}
\tablehead{\colhead{Visual} & \multicolumn{2}{c}{Total} &
  \multicolumn{2}{c}{Complex (\%)\tablenotemark{a}} &
  \multicolumn{2}{c}{Resolved (\%)\tablenotemark{a}} & 
  \multicolumn{2}{c}{Compact (\%)\tablenotemark{a}} \\
\colhead{Classification} &
\colhead{\emph{QSOs}} & \colhead{\emph{Gals.}} &
\colhead{\emph{QSOs}} & \colhead{\emph{Gals.}} &
\colhead{\emph{QSOs}} & \colhead{\emph{Gals.}} &
\colhead{\emph{QSOs}} & \colhead{\emph{Gals.}} } 
\startdata
 Complex & 202 & 286 & 77 & 95 &  8 &  4 & 15 &  1 \\
Resolved &  87 &  79 &  8 & 42 & 72 & 58 & 20 &  0 \\
 Compact & 273 & 144 &  5 & 26 &  9 & 26 & 86 & 48 \\
\enddata
\tablecomments{Results compiled for a set of 1071
radio quasars and radio galaxies (identified spectroscopically from the SDSS).
Visual classification is listed at the left; automatic classification is listed
along the top.}
\tablenotetext{a}{Numbers refer to the percentage of quasars/galaxies in each visual
  classification category which received a particular automatic classification.}
\end{deluxetable}

\begin{deluxetable}{llrrrr}
\tablewidth{0in}
\tablecaption{\label{table:alphamean}
Gaussian fits to $(\alpha^{92}_{20}+\alpha^{20}_6)/2$ distributions}
\tablehead{\colhead{Subclass} & \colhead{$\deg^{-2}$} &
  \colhead{mean\tablenotemark{a}} & \colhead{median} &
  \colhead{FWHM\tablenotemark{a}}} 
\startdata
\sidehead{Complex\tablenotemark{b}}
All & 1.3 & -0.868 & -0.863 & 0.433 \\
no SDSS & 1.1 & -0.886 & -0.881 & 0.406 \\
galaxy\tablenotemark{c} & 0.15 & -0.742 & -0.732 & 0.433 \\
pt. source\tablenotemark{d} & 0.069 & -0.705 & -0.662 & 0.629 \\
\\
\sidehead{Resolved\tablenotemark{b}}
All & 0.68 & -0.850 & -0.843 & 0.433 \\
no SDSS & 0.38 & -0.895 & -0.892 & 0.416 \\
galaxy\tablenotemark{c} & 0.19 & -0.808 & -0.805 & 0.385 \\
pt. source\tablenotemark{d} & 0.11 & -0.742 & -0.733 & 0.571 \\
\\
\sidehead{Compact\tablenotemark{b}}
All & 0.99 & -0.713 & -0.680 & 0.822 \\
no SDSS & 0.45 & -0.839 & -0.815 & 0.530 \\
galaxy\tablenotemark{c} & 0.24 & -0.710 & -0.670 & 0.682 \\
pt. source\tablenotemark{d} & 0.30 & -0.302 & -0.288 & 1.03 \\
\enddata
\tablecomments{Values correspond to histograms shown in 
Figure~\ref{fig:alphamean all radio}.  Sources were detected by all four radio
surveys, and follow the additional constraint $t_\mathrm{NVSS}<12$.}
\tablenotetext{a}{Estimated from a Gaussian fit to the
  $(\alpha^{92}_{20}+\alpha^{20}_6)/2$ distribution.} 
\tablenotetext{b}{20\,cm radio morphology.}
\tablenotetext{c}{Unresolved by SDSS.}
\tablenotetext{d}{Resolved by SDSS.}
\end{deluxetable}

\begin{deluxetable}{lrrrr}
\tablewidth{0in}
\tablecaption{\label{table:alpha9220 gb6}
Gaussian fits to $\alpha^{92}_{20}$ distributions}
\tablehead{\colhead{Subclass} & \colhead{$\deg^{-2}$} &
  \colhead{mean\tablenotemark{a}} & \colhead{median} &
  \colhead{FWHM\tablenotemark{a}}} 
\startdata
\sidehead{Complex\tablenotemark{b}}
All & 1.3 & -0.786 & -0.782 & 0.323 \\
no SDSS & 1.1 & -0.798 & -0.797 & 0.301 \\
galaxy\tablenotemark{c} & 0.15 & -0.659 & -0.654 & 0.323 \\
pt. source\tablenotemark{d} & 0.069 & -0.718 & -0.690 & 0.396 \\
\\
\sidehead{Resolved\tablenotemark{b}}
All & 0.68 & -0.764 & -0.765 & 0.339 \\
no SDSS & 0.38 & -0.807 & -0.817 & 0.320 \\
galaxy\tablenotemark{c} & 0.19 & -0.700 & -0.702 & 0.311 \\
pt. source\tablenotemark{d} & 0.11 & -0.713 & -0.710 & 0.378 \\
\\
\sidehead{Compact\tablenotemark{b}}
All & 0.99 & -0.632 & -0.563 & 0.640 \\
no SDSS & 0.45 & -0.727 & -0.683 & 0.440 \\
galaxy\tablenotemark{c} & 0.24 & -0.593 & -0.553 & 0.576 \\
pt. source\tablenotemark{d} & 0.30 & -0.271 & -0.229 & 1.05 \\
\enddata
\tablecomments{Values correspond to the data samples investigated in
  Fig.~\ref{fig:alphamean all radio} and Table~\ref{table:alphamean}.  Sources
  were detected by all four radio surveys, and follow the additional constraint
  $t_\mathrm{NVSS}<12$.  The $\alpha^{92}_{20}$ histograms are not shown.}
\tablenotetext{a}{Estimated from a Gaussian fit to the distribution.}
\tablenotetext{b}{20\,cm radio morphology.}
\tablenotetext{c}{Unresolved by SDSS.}
\tablenotetext{d}{Resolved by SDSS.}
\end{deluxetable}

\clearpage

\begin{deluxetable}{lrrrccc} 
\tablewidth{0in}
\tablecaption{\label{table:radiostars}
Candidate radio star selection}
\tablehead{ 
\colhead{Data Source} & \colhead{All\tablenotemark{a}} & \colhead{in
  FIRST\tablenotemark{b}} & \colhead{$i < 19.0$\tablenotemark{c}} & 
\colhead{Non-QSO colors\tablenotemark{d}} & \colhead{w/ spec.\tablenotemark{e}} 
& \colhead{stellar spec.\tablenotemark{f}} } 
\startdata 
opt. point sources\tablenotemark{e}
& 19,133,672 &  12,100 &  2,424  &  532  &  406  & 78 \\ 
quasar catalog\tablenotemark{g}
&    35,450  &  2,927  &  1,761  &  289  &  289  &  2 \\
&   (37,010) & (3,147) & (1,931) & (358) & (358) & (2) \\
\enddata 
\tablecomments{}
\tablenotetext{a}{All sources in the catalog overlap region.  (For the optical
  point source sample, there are 5,841,685 sources with $i<19$).}
\tablenotetext{b}{Within 1 arcsec from a FIRST source.}
\tablenotetext{c}{Using model magnitudes (1,653 for quasar catalog when using
  psf magnitudes.)} 
\tablenotetext{d}{Selected by rejecting sources with
  $u-g<0.8$ AND $-0.2 < g-r < 0.6$.} 
\tablenotetext{e}{Selected from SDSS Data Release 6.}
\tablenotetext{f}{Visually-inspected to confirm automatic classification as
  stellar spectrum.} 
\tablenotetext{g}{Schneider et al. (2006) catalog corresponding to an
  unresolved SDSS source (values in parentheses correspond to the entire quasar
  catalog).} 
\end{deluxetable}

\clearpage

\begin{deluxetable}{clcrr}
\tablecolumns{5}
\tablewidth{0in}
\tablecaption{\label{table:data}
Catalog subsets available for download}
\tablehead{
\colhead{Sample} & \colhead{Description} & \colhead{Selection} &
\colhead{\# rows}  & \colhead{size\tablenotemark{a}} }
\startdata 
A & Complete catalog & None & 2,724,343 & 257M (1.4G) \\
\\
B & $100\deg^2$ subset & $140\degr<$\emph{ra}$<165\degr$ & 16,453 & 3.4M (15M) \\
  & & $40\degr<$\emph{dec}$<50\degr$ \\
\\
C & Detected by FIRST & \emph{overlap}\,=\,1 & 141,881 & 5.4M (8.7M) \\
  & and NVSS & \emph{matchflag\_nvss}\,$=-1$ \\
  & & \emph{matchflag\_first}\,=\,1 \\
  & & \emph{distance}\,$\leq25\arcsec$ \\
\\
D & Detected by FIRST, & (subset of C) & 63,660 & - \\
  & NVSS, and WENSS & \emph{wenss\_flux}\,$\neq0$ \\
  & & \emph{wenss\_distance}\,$\leq30\arcsec$ \\
\\
E & Detected by FIRST, & (subset of D) & 12,414 & - \\
  & NVSS, WENSS, & \emph{gb6\_flux}\,$\neq0$ \\
  & and GB6 & \emph{gb6\_distance}\,$\leq70\arcsec$ \\
\\
F & Detected by FIRST, & (subset of E)\tablenotemark{b} & 4,732 & 625K \\
  & NVSS, WENSS, GB6, & \emph{near\_type}\,$\neq0$ \\
  & and SDSS& \emph{near\_distance}\,$\leq2\arcsec$ \\
\\
G & Spectroscopic galaxies & (subset of D)\tablenotemark{b} & 2,885 & 439K \\
  & detected by FIRST, & \emph{spec\_type}\,$=2$ \\
  & NVSS, WENSS, & \emph{near\_distance}\,$\leq2\arcsec$ \\
  & and SDSS \\
\\
H & Spectroscopic quasars & (subset of C)\tablenotemark{c} & 1,288 & 755K \\
  & detected by FIRST,& Match to Schneider et \\
  & NVSS, WENSS, SDSS & al. (2007) within $2\arcsec$ \\ 
\\
I & Isolated\tablenotemark{d} FIRST-NVSS & (subset of C) & 109,825 & 4.1M (6.8M) \\
  & & \emph{distance}\,$\leq15\arcsec$ \\
  & & No other FIRST \\
  & & sources within $30\arcsec$ \\
\\
J & Isolated\tablenotemark{d} FIRST-NVSS, & (subset of I)\tablenotemark{b} & 44,851 & 3.7M (5.7M) \\
  & and SDSS & \emph{near\_type}\,$\neq0$ \\
   & & \emph{near\_distance}\,$\leq2\arcsec$ \\
\\
K & Candidate high- & (subset of E) & 9,619 & 2.5M (5.1M) \\
  & redshift galaxies & (\emph{near\_ra}=0 OR \\
  & ($z\gtrsim1$) & \emph{near\_distance}\,$>3\arcsec$) \\
  & & $\alpha^{92}_{20}\leq-0.5,\,\alpha^{20}_{6}\leq-0.5$ \\
  & & remove suspicious \\
  & & objects (\S\ref{subsec:highz gals}) \\
\enddata 
\tablecomments{The first column gives the subset identifier, the second column
contains a brief description, the third column lists the requirements to
select the data set from the complete catalog, and the fourth column shows the
number of sources in the data set.  The fifth column indicates the size of the
downloadable file for datasets available (with limited data parameters) on the
catalog website.  For compressed files, the uncompressed size is listed in
parentheses.}
\tablenotetext{a}{File size.  For compressed files, number in parentheses
  refers to uncompressed size.  For samples which are subsets of downloadable
  files (in terms of both selected objects and relevant data parameters), no
  size is given.}
\tablenotetext{b}{While these objects are a subset of another sample, the file
  includes additional SDSS data parameters.}
\tablenotetext{c}{The sample of spectroscopic quasars is a subset of objects in
  Sample~C; however the file includes data parameters from the
  \citet{dr5quasars} catalog which are not part of the unified radio catalog.} 
\tablenotetext{d}{Isolated in this context refers to FIRST-NVSS pairs with no
  other FIRST sources within $30\arcsec$.}
\end{deluxetable}

\end{document}